\documentclass{article}

\usepackage{arxiv}

\usepackage[utf8]{inputenc} 
\usepackage[T1]{fontenc}    
\usepackage{hyperref}       
\usepackage{url}            
\usepackage{booktabs}       
\usepackage{amsfonts}       
\usepackage{nicefrac}       
\usepackage{microtype}      
\usepackage{lipsum}		
\usepackage{graphicx}
\usepackage{doi}
\usepackage{amsmath}
\usepackage{graphicx}
\usepackage{booktabs,tabularx,makecell}
\usepackage{ragged2e} 
\newcolumntype{Y}{>{\RaggedRight\arraybackslash\tiny\hsize=\hsize}X}

\usepackage[
  backend=biber,
  style=apa,
  natbib=true    
]{biblatex}
\title{\emph{Self++}: Co-Determined Agency for Human--AI Symbiosis in Extended Reality}

\author{ \href{https://orcid.org/0000-0001-5870-9221}{\includegraphics[scale=0.06]{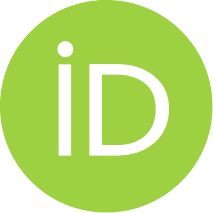}\hspace{1mm}Thammathip Piumsomboon}\thanks{https://www.quadriclab.org/} \\
	School of Product Design\\
	University of Canterbury\\
	Christchurch, New Zealand \\
	\texttt{tham.piumsomboon@canterbury.ac.nz} \\
}

\renewcommand{\shorttitle}{\textit{Self++: Co-Determined Agency for Human--AI Symbiosis in XR}}

\hypersetup{
pdftitle={Self++},
pdfsubject={HAI},
pdfauthor={Thammathip Piumsomboon},
pdfkeywords={Extended Reality (XR), Human-AI Symbiosis, Human-AI Teaming, Self-Determination Theory, Free Energy Principle, Co-Determination, Augmented Agency, Co-determined Agency, Interaction Design, Trustworthy AI},
}

\begin{document}
\maketitle

\begin{abstract}
    Self++ is a design blueprint for human–AI symbiosis in extended reality (XR) that preserves human authorship while still benefiting from increasingly capable AI agents. Because XR can shape both perceptual evidence and action, apparently ‘helpful’ assistance can drift into over-reliance, covert persuasion, and blurred responsibility. Self++ grounds interaction in two complementary theories: Self-Determination Theory (autonomy, competence, relatedness) and the Free Energy Principle (predictive stability under uncertainty). It operationalises these foundations through co-determination, treating the human and the AI as a coupled system that must keep intent and limits legible, tune support over time, and preserve the user’s right to endorse, contest, and override. These requirements are summarised as the co-determination principles (T.A.N.): Transparency, Adaptivity, and Negotiability. Self++ organises augmentation into three concurrently activatable overlays spanning sensorimotor competence support (Self: competence overlay), deliberative autonomy support (Self+: autonomy overlay), and social and long-horizon relatedness and purpose support (Self++: relatedness and purpose overlay). Across the overlays, it specifies nine role patterns (Tutor, Skill Builder, Coach; Choice Architect, Advisor, Agentic Worker; Contextual Interpreter, Social Facilitator, Purpose Amplifier) that can be implemented as interaction patterns, not personas. The contribution is a role-based map for designing and evaluating XR-AI systems that grow capability without replacing judgment, enabling symbiotic agency in work, learning, and social life and resilient human development.
\end{abstract}


\begin{figure}[t!]
\centering
    \includegraphics[width=0.7\textwidth]{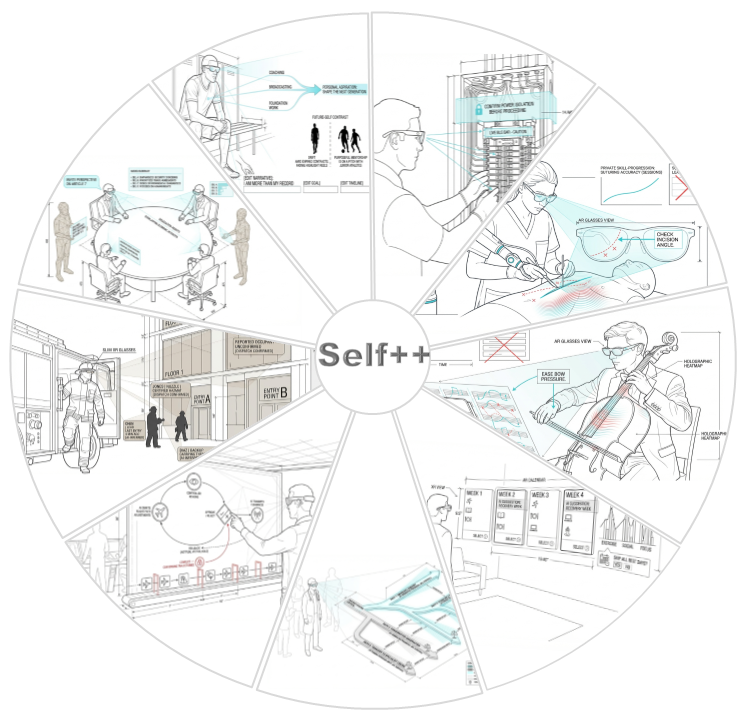}
    \caption{The nine Self++ role patterns organised across three concurrently activatable overlays, with co-determination principles (T.A.N.) scaling in strength with overlay scope and initiative. \textbf{Overlay~1 (Self): Competence support.} R1~--- \emph{Tutor}: reduces novice uncertainty through a safe, learnable corridor (e.g., a trainee electrician receives anchored directional arrows, step gating, and ghosted hand exemplars through XR glasses while working on a residential electrical panel). R2~--- \emph{Skill Builder}: calibrates and generalises skill through variability and augmented feedback (e.g., a training doctor receives real-time motion traces and a holographic accuracy heatmap overlaid onto a practice mannequin during a surgical procedure). R3~--- \emph{Coach}: builds robustness under stress and supports self-correction (e.g., a cellist receives intonation feedback, fingerboard pressure heatmaps, and metacognitive prompts during a live performance, with social comparison replaced by private progression tracking). \textbf{Overlay~2 (Self+): Autonomy support.} R4~--- \emph{Choice Architect}: shapes the decision context while preserving authorship (e.g., a person views a floating AR monthly calendar where recovery weeks are gently highlighted and a friction gate requests confirmation before overriding rest days). R5~--- \emph{Advisor}: externalises deliberation by making counterfactuals and trade-offs inspectable (e.g., an ER doctor sees a branching holographic decision tree with uncertainty bands, survival-confidence estimates, and provenance badges distinguishing AI prognosis from attending physician input). R6~--- \emph{Agentic Worker}: executes delegated tasks under a proposal-approval loop with rollback (e.g., an air traffic control shift manager oversees an AI-drafted routing queue where conflict items are flagged and rerouted back for manual handling, with any clearance reversible before transmission). \textbf{Overlay~3 (Self++): Relatedness and purpose support.} R7~--- \emph{Contextual Interpreter}: makes identity, norms, and downstream impacts legible to reduce social surprise (e.g., a firefighter arriving at an incident sees AR-labelled crew roles, building entry points, and provenance badges distinguishing dispatch-confirmed from AI-inferred information). R8~--- \emph{Social Facilitator}: improves human-to-human coordination and repair (e.g., diplomats at a round-table negotiation receive personalised, opt-in AR overlays including speaking-time balance, perspective-invitation prompts, and neutral micro-summaries of each delegation's position, while embodied virtual agents surface shared precedents as common ground). R9~--- \emph{Purpose Amplifier}: supports long-horizon value coherence by making future trajectories legible and editable (e.g., a retiring athlete views a holographic value-map converging personal strengths toward an aspiration, with a future-self contrast between drift and purposeful mentorship and editable identity-narrative fields).}
    \label{fig:selfpp_roles}
\end{figure}

\section{Introduction}

Seen in a longer arc, the present moment is a new chapter in human cognitive extension. Early visions of human--computer integration already anticipated a deep mind--machine symbiosis. In 1960, J. C. R. Licklider described "man-computer symbiosis" as an interactive partnership where humans and computers complement each other’s strengths \citep{licklider2008man}. Shortly after (1962), Douglas Engelbart proposed augmenting human intellect to boost problem-solving through technology \citep{engelbart2021augmenting}. Later theories formalised this relationship: the extended mind hypothesis argued that artifacts can become literal parts of cognition \citep{clark1998ExtendedMind}, while distributed cognition emphasised that thinking often spans people, tools, and environments \citep{hutchins1995cognition}. Empirical work supports parts of longstanding worries about cognitive erosion, including the "Google effect" \citep{sparrow2011google}, inflated perceived knowledge from internet access \citep{fisher2015searching}, and weakened spatial memory from heavy GPS reliance \citep{dahmani2020habitual}. At the same time, cognitive science and HCI stress that "thinking with things" can improve reasoning \citep{kirsh2010thinking}, and that distributing cognitive load can enable more complex problem-solving \citep{hollan2000distributed}.

We are accelerating toward more autonomous productivity, driven by virtual agents and embodied robots. Toolchains, platforms, and organisational agent stacks are being optimised for throughput, reliability, and delegation, first under human instruction and, plausibly, under higher-level supervisory agents. This trajectory sharpens an old question: how do we gain the benefits of delegation without losing authorship? Much current work focuses on catastrophic risks, from extreme harm like misalignment to subtler manipulation and societal subordination \citep{Bostrom2014Superintelligence, bostrom2018ethics}. Alongside these frontier concerns is a nearer failure mode: AI-mediated convenience can erode careful reasoning, making people less critical and more easily persuaded \citep{lee2025impact}. This predates modern generative AI, but today’s systems are more fluent, more personalised, and more persuasive at scale.

This acceleration represents a major expansion of the human cognitive niche. Evolutionary accounts suggest humans gained advantage through causal reasoning, tool-making, and cooperative action rather than biological arms races \citep{pinker2010cognitive}. Humans externalised thinking into artifacts and social systems, expanding intelligence through culture. This tendency is captured by the notion that we are "natural-born cyborgs" \citep{Clark2003NaturalCyborgs} living in co-evolution with our technologies \citep{Clark2008Supersizing}. Today, the niche expands again through Extended Reality (XR) and Artificial Intelligence (AI). XR extends perception and presence in mixed and virtual environments, while AI externalises reasoning by perceiving and acting alongside us. The result is a tightly interwoven human--machine ecology in which cognitive strategies are co-generated.

This expansion also carries costs. The complexity and volatility of an XR--AI ecosystem can create an ``entropy challenge'': unpredictable stimuli and shifting agent behaviours that outstrip our capacity to maintain coherent models \citep{Friston2010FreeEnergy, Clark2008Supersizing}. The mismatch can manifest as information overload, attentional fragmentation, and blurred agency where users may be unsure where their intentions end and the system’s suggestions begin. These symptoms point to design failures in how autonomy, competence, and social connection are protected under delegation. Emerging phenomena make this gap visible, including split-attention demands in XR workflows \citep{Thanyadit2022XRLive}, identity confusion as agents become more human-like \citep{Zhang2024Triplets, zhang2025hat}, and misaligned persuasion where simulation success does not translate into real behaviour change \citep{doudkin2025synthetichumangap}. An ``AI loneliness trap'' may also emerge, where convenient synthetic companionship gradually displaces human relationships for some users \citep{liu2025heterogeneous}.

If AI is the latest extension of cognition, synergy is not automatic. Automation research has long catalogued challenges such as common ground, trust calibration, and complacency \citep{klein2004teamplayer}. A large meta-analysis suggests that human--AI teams often fail to outperform the better of the human or AI alone \citep{vaccaro2024combinations, yousefi2025team}, implying that effective teaming must be deliberately designed with interaction mechanisms that respect human cognition and shared agency. Recent work reinforces that the most accurate AI is not always the best teammate \citep{bansal2021most}. Human--computer integration research argues for a deeper intertwining where systems adapt to the user’s state in real time and users fluidly delegate and reclaim control \citep{mueller2023toward}. XR makes this tangible by placing intelligent support in the user’s perceptual and social space. AR enables in-situ augmentation \citep{zhou2008trends, kim2018revisiting}, including social interventions like calming virtual companions \citep{norouzi2022advantages}, while VR provides rich environments for studying collaboration dynamics \citep{pan2018and, Wen2025GenLinguaScape}. Together, these platforms make the hybrid niche observable and designable, strengthening the case for co-determined agency as a design primitive.

Self++ responds to this agency problem by grounding design in two complementary foundations: Self-Determination Theory (SDT; autonomy, competence, relatedness) and the Free Energy Principle (FEP; predictive stability under uncertainty). In simple terms, SDT specifies what must be preserved for flourishing, while FEP explains why the pressure intensifies as environments become more volatile and mediated. We operationalise these foundations through \emph{co-determination}: treating human and AI as a coupled system that makes intent and limits legible, tunes support to development, and preserves the user’s right to endorse, contest, and override assistance. We summarise these requirements as three co-determination principles (T.A.N.): Transparency, Adaptivity, and Negotiability.

These pressures matter deeply to me as an educator and parent. Self++ is not meant to diminish AI’s utility or reject autonomy as a design direction. Instead, it offers a framework for human--AI systems that preserve the benefits of intelligent support while strengthening user agency, competence, and social integration. Self++ organises augmented agency into three concurrently activatable overlays and nine role patterns spanning sensorimotor support, deliberative decision support, and longer-horizon social and purpose alignment. Across all role patterns, T.A.N. functions as a practical safeguard so uncertainty regulation supports human development rather than bypassing it.

\section{Background}

\subsection{Theoretical Foundations: Self-Determination and Free Energy}

As noted in the Introduction, the emerging XR--AI ecology raises the stakes for supportive AI design. Self++ is grounded in two complementary pillars: Self-Determination Theory (SDT) \citep{Ryan2000SDT} and the Free Energy Principle (FEP) \citep{Friston2010FreeEnergy}. Together, they explain why autonomy, competence, and relatedness matter under intelligent mediation, and how systems might support them without eroding agency. Self-Determination Theory identifies three basic psychological needs (autonomy, competence, and relatedness) as essential for motivation and well-being \citep{Ryan2000SDT}. Autonomy is volitional, self-endorsed action; competence is efficacy and skill growth; relatedness is connection and belonging. When these needs are supported, people show stronger intrinsic motivation, learning, and well-being; when they are chronically frustrated, disengagement and poorer performance follow.

The Free Energy Principle complements SDT by explaining why these needs become harder to protect as environments grow more complex. Predictive-processing accounts formalise perception and action as approximate Bayesian inference aimed at minimising surprise \citep{clark2013whatever}. Friston’s FEP generalises this: biological systems maintain integrity by minimising free energy, closely related to long-run prediction error \citep{Friston2010FreeEnergy}. The brain updates internal models to keep sensory input within expected bounds; as environments become more volatile, the computational and attentional cost of this stabilisation rises. Because human predictive models evolved for relatively stable ecologies, they can be mismatched to fast-changing, algorithmically mediated XR, producing sustained prediction errors experienced as stress, confusion, or overload \citep{Clark2008Supersizing}.

Self++ bridges SDT and FEP by treating autonomy, competence, and relatedness as stabilising conditions for predictive cognition: competence improves anticipatory control, autonomy protects against externally imposed goals that conflict with internal priors, and relatedness offloads uncertainty onto shared social models.

To operationalise these needs in interaction design, we propose \textit{co-determination}. Rather than command-following tools or unilateral autonomy, co-determination treats human and AI as a coupled system that negotiates control to preserve psychological stability and minimise prediction error. We summarise this stance as three co-determination principles (\textit{T.A.N.}):

\begin{itemize}
    \item \textbf{Transparency:} From an FEP perspective, the system should minimise ``hidden states.'' If an agent’s reasoning or uncertainty is opaque, users cannot predict its behaviour, increasing anxiety and miscalibrated trust. Transparency supports accurate user mental models \citep{clark2013whatever}.
    \item \textbf{Adaptivity:} From an SDT perspective, support should track the user’s changing \textit{competence}. Static assistance becomes either intrusive (thwarting autonomy) or insufficient (thwarting competence). Adaptivity provides scaffolding that intensifies or fades to match learning and context \citep{vygotsky1978mind}.
    \item \textbf{Negotiability:} Rooted in \textit{autonomy}, users must be able to endorse, decline, or override system actions. Without negotiation and reversal, users lose authorship and volition \citep{Ryan2000SDT}.
\end{itemize}


\subsection{Philosophical and Cultural Influence}

SDT and FEP explain how humans sustain motivation and stability under uncertainty, but they leave a prior normative question open: what kinds of selves are being shaped as perception, action, and social interaction become increasingly mediated by intelligent systems? This subsection adds a philosophical and cultural framing for why agency, interpretation, and social embeddedness must remain central and why T.A.N. becomes an ethical requirement.

Across many traditions, selfhood is relational and enacted through conditions rather than fixed or self-contained. In Buddhist philosophy, the self is an impermanent process (anattā) arising through interdependence \citep{gallagher2024self}. Māori ontology similarly foregrounds relational identity through whakapapa and frames well-being as sustained through balanced relationships among people, communities, and environment \citep{wilson2021creating}. Cross-cultural psychology likewise distinguishes relational and individual-centred models of self, showing that meaning, obligation, and autonomy are socially situated \citep{markus2014culture}. These views do not reject autonomy; they recast it as accountable and context-sensitive.

Cognitive science offers a parallel account. Embodied and enactive approaches argue that sense-making emerges through ongoing organism–environment coupling, not detached internal reconstruction \citep{varela2017embodied, gallagher2006body, di2010horizons}. The extended mind thesis similarly holds that cognition can span brain, body, artefacts, and social structures \citep{clark1998ExtendedMind}. If selves are enacted through tools and relationships, AI is not just an external utility; it helps shape the conditions through which identity and experience are repeatedly constructed. Self++, therefore, treats the social environment as a functional substrate for agency and understanding, not an optional dimension.

A further implication of this relational stance is that what counts as autonomy, competence, and relatedness, and how they are weighted, is not culturally uniform. Cross-cultural psychology has long shown that the boundaries of the autonomous self, the role of obligation in competent action, and the forms of belonging that sustain well-being differ substantially across individual-centred and relational self-construals \citep{markus2014culture}. Indigenous frameworks such as Māori models of relational health foreground collective accountability and intergenerational connection as conditions for flourishing, not merely as a context for individual choice \citep{wilson2021creating}. Buddhist accounts treat selfhood as processual and interdependent, making the very notion of a bounded "autonomous agent" a convention rather than a ground truth \citep{gallagher2024self}. These differences are not peripheral to Self++; they are a primary reason why the framework adopts a procedural rather than substantive ethical stance. Self++ does not prescribe which values are correct or which model of selfhood is authoritative. Instead, it specifies interactional conditions (transparency, adaptivity, and negotiability) under which users and communities can recognise, reflect on, and act from their own endorsed commitments, whatever those commitments may be. T.A.N. is designed for moral and cultural plurality: transparency makes influence visible so it can be evaluated against local norms; adaptivity prevents the system from freezing around a single cultural default; and negotiability gives individuals and groups the power to contest, reconfigure, or refuse what the system surfaces and how. This procedural stance does not claim neutrality, choosing what to make legible is itself a normative act, but it does ensure that such choices remain inspectable and revisable rather than hard-coded.

This relational stance is especially relevant for AI design when read alongside Dependent Origination (pratītyasamutpāda) and predictive processing. Dependent Origination holds that experience arises through interdependent causes and conditions. Predictive processing makes a structurally similar claim: perception is generated by hierarchical generative models that predict sensory input, so experience depends on the negotiation between signals and learned expectations \citep{Friston2010FreeEnergy, clark2013whatever, hohwy2013predictive}. Prior work notes resonances between Buddhist accounts of interdependent experience and predictive approaches \citep{ho2022intersubjectivity}. Building on this, we map Dependent Origination to inferential perception: ignorance reflects model mismatch; formations shape priors; sense bases/contact sample evidence; and craving/clinging reflects the drive to resolve uncertainty, potentially hardening priors. The upshot is that perceptual qualities (e.g., a virtual object’s apparent “redness”), or qualia, are not intrinsic to stimuli, but emerge from relational conditions spanning input and interpretation.

A key ethical implication follows: if mediation conditions experience and identity, those conditions must be legible and adjustable. Predictive accounts and the FEP formalise this dependency: perception and action reflect ongoing model--evidence negotiation, and uncertainty minimisation can become maladaptive when systems push users toward premature closure, rigid priors, or habitual over-reliance \citep{friston2017active, parr2022active}. The design question is therefore normative, not merely technical: XR and AI reshape the informational and social conditions that guide interpretation, attention, and obligation, thereby shaping the self that is enacted over time. SDT specifies acceptable direction for this influence: assistance should expand autonomy (authorship), build competence (capability, not substitution), and sustain relatedness (trust and belonging) \citep{Shneiderman2020HCAI, capel2023human}.

The interactional requirement of co-determination is what makes such conditioning ethically tractable, and T.A.N. can be read as co-determination principles for an ethical “extended self”:

\begin{itemize}
    \item \textbf{Transparency as Insight:} Because mediation shapes experience, intent, bias, and uncertainty must be visible. Transparency clarifies what is being altered, why, and with what confidence, enabling trust calibration \citep{clark2013whatever}.
    \item \textbf{Adaptivity as Impermanence:} Users develop; support must change with them. Adaptivity tunes (and fades) assistance as goals, context, and competence evolve, avoiding stale assumptions and dependency \citep{amershi2019guidelines}.
    \item \textbf{Negotiability as Volitional Action:} Users must be able to consent, contest, and override. Without meaningful veto and reversal, systems displace authorship and moral responsibility \citep{Shneiderman2020HCAI, mueller2023toward}.
\end{itemize}

Together, this framing shows why T.A.N. is an ethical requirement, not merely a “trust mechanism”: mediated systems must be transparent, adaptive, and negotiable so uncertainty is regulated with users rather than for them.

The ethics of manipulation literature provides a complementary negative argument for these requirements. Philosophical accounts identify three main characterisations of manipulative influence: bypassing rational deliberation, trickery (inducing faulty mental states), and pressure (non-coercive but difficult-to-resist influence) \citep{noggle2022manipulation}. Manipulation is widely held to undermine the validity of consent \citep{faden1986history} and to ``pervert the way that person reaches decisions, forms preferences, or adopts goals'' \citep{raz1988morality}. More recent work characterises it as a hidden influence that targets cannot easily become aware of \citep{susser2019technology}. These characterisations are directly relevant to XR--AI systems, where perceptual mediation, personalised nudging, and delegated action all create conditions under which influence can become covert, difficult to resist, or substitutive of the user's own reasoning. T.A.N. can be read as a systematic defence against all three forms. Transparency prevents trickery by making the system's intent, reasoning, and uncertainty legible, so users cannot be induced into faulty beliefs about what is being influenced or why. Negotiability prevents pressure by ensuring the user always retains a viable exit; consent, override, and revocability eliminate the ``awkward and difficult to resist'' condition that characterises manipulative pressure \citep{faden1986history}. Adaptivity prevents the subtlest form, bypassing rational deliberation, by ensuring that support engages and progressively strengthens the user's own deliberative capacities rather than substituting for them; a system that never fades its scaffolding effectively outsources rational deliberation, and over time, such functional outsourcing becomes indistinguishable from bypassing it. Under these conditions, the system's influence operates not by bypassing or subverting rational deliberation, but by scaffolding it, providing the informational and attentional conditions under which the user can deliberate more effectively while retaining full authorship over the resulting decision. In this reading, co-determination is not merely non-manipulative influence; it is structurally anti-manipulative, because it preserves and strengthens the very deliberative processes that manipulation subverts.


\subsection{Extended Reality as a Perceptual Filter: Dependent Origination and Predictive Control}

Extended Reality (XR), encompassing Augmented Reality (AR), Mixed Reality (MR), and Virtual Reality (VR), is a technological frontier for modulating human perception. Recent work frames XR systems as technologies that can modulate the incoming light field itself rather than merely overlay virtual content, highlighting that XR operates as a perceptual filter acting prior to conscious interpretation \citep{rendon2025augmented}. Beyond addition, XR enables the subtraction or alteration of sensory input through techniques such as Diminished Reality \citep{mori2017survey}, allowing aspects of the environment to be suppressed or transformed. Together, these capabilities constitute a form of mediated reality in which XR actively filters perceptual evidence rather than passively displaying information. By selectively amplifying, attenuating, or removing stimuli, XR systems shape what users attend to and how they interpret their surroundings. Perceptual filtering should therefore not be treated as a neutral presentation choice, but as an intervention that warrants disclosure of what is being altered and a user-legible rationale for why. This perspective also motivates XR as a systematic testbed for human--AI interaction research. \citep{wienrich2021extended} propose an XR--AI continuum and ``eXtended AI,'' arguing that XR can be used to prototype and study prospective AI embodiments and interfaces in controlled, high-fidelity contexts before deployment.

In practice, XR-mediated filtering can be applied in both constructive and protective ways. Constructively, XR can foreground task-relevant cues or reveal otherwise inaccessible information, supporting learning and decision-making \citep{zollmann2020visualization, Dong2024XRDecision}. Altered viewpoints and scale transformations can also enable users to reason about spatial or structural relationships beyond ordinary physical constraints \citep{piumsomboon2018mini, Piumsomboon2018super, Piumsomboon2019Giant}. Protectively, XR can suppress distracting or intrusive stimuli, such as visually cluttered backgrounds or advertising content, to reduce attentional fragmentation and cognitive load \citep{Katins2025AdBlockedXR}. These interventions reconfigure the informational structure of the environment to better align with task demands and user capacity, but they also introduce a distinctive risk: because XR can reshape “ground truth” at the level of perception, the possibility of false beliefs, miscalibrated trust, or socially consequential omission becomes acute. This motivates treating the co-determination principles (T.A.N.) not as abstract ethical commitments, but as concrete constraints for responsible perceptual regulation.

\begin{itemize}
    \item \textbf{Transparency in Perception:} The system must disclose how it is filtering reality. If an XR system suppresses visual noise (e.g., removing ads or clutter \citep{Katins2025AdBlockedXR}), users must be aware that information is being hidden and why. Perceptual transparency prevents users from mistaking a curated evidential stream for objective reality.
    \item \textbf{Adaptivity in Scaffolding:} Perceptual enhancements, such as highlighting task-relevant cues \citep{zollmann2020visualization}, should not become permanent crutches. True adaptivity implies that as a user learns to notice patterns (increasing perceptual competence), highlights can fade or re-target, transferring predictive load back to the user while preserving support when conditions change.
    \item \textbf{Negotiability of Reality:} Users must have the power to define and revise their perceptual boundaries. Whether it is a therapeutic application modulating anxiety triggers or exposure in PTSD treatment \citep{rizzo2014virtual}, or a productivity tool filtering distractions, users should be able to inspect, override, and revert filtering on demand, including simple “show me what was removed” controls.
\end{itemize}

First, XR can function as a perceptual enhancer to reduce surprise by providing timely, task-relevant cues that make situations more predictable. If perception is, as Andy Clark suggests, a kind of “controlled hallucination” constrained by sensory feedback, then XR can be understood as an externalised intervention in the evidence that stabilises perceptual inference \citep{Wiese2025ConsciousAR}. By enhancing signal quality or suppressing noise, XR systems shape the feedback that constrains expectations, making environments more predictable and cognitively manageable. Examples include AR navigation overlays that reduce wayfinding ambiguity \citep{zollmann2020visualization}, military helmet displays that stabilise situational awareness in fast-changing environments \citep{livingston2011military}, and surgical AR systems that integrate imaging data directly into the operative field \citep{sielhorst2008advanced}. However, cueing can also become a permanent “perceptual crutch” if it is not designed to fade as competence develops. Adaptivity, therefore, implies scaffolding that can be intensified, faded, or re-targeted as user skill and context change, transferring predictive load back to the user as they learn to notice relevant patterns.

Conversely, XR can be used as a perceptual filter to minimise surprise or stress by attenuating extraneous or harmful inputs. The Ad-Blocked Reality prototype, for instance, digitally obscured billboard advertisements and reported increased focus and subjective calm while raising concerns about the removal of potentially relevant information \citep{Katins2025AdBlockedXR}. More broadly, XR interventions have been characterised as perceptual manipulations that can extend beyond immediate action to influence longer-term cognitive processes, including memory and self-perception \citep{Bonnail2023Memory}. In therapeutic contexts, XR-based filtering has also been proposed as a means of attenuating anxiety triggers or modulating exposure in PTSD treatment, enabling gradual recalibration of expectations under controlled conditions \citep{rizzo2014virtual}. This illustrates a central trade-off: while filtering can enhance immediate well-being and a sense of control, excessive or opaque filtering risks distorting the user’s world model and narrowing attention in ways analogous to perceptual echo chambers. Negotiability is therefore not optional. Users should be able to inspect, override, and revert filtering on demand, including simple “show me what was removed” controls, especially where safety, fairness, or civic relevance are at stake.

These ethical requirements become even clearer in fully synthetic VR. By presenting a largely constructed sensorium, VR allows systematic manipulation of the relationship between expectation and sensation, making the role of prior beliefs in shaping experience explicit. Classic embodiment illusions, such as virtual limb or full-body ownership, arise when visual and sensorimotor contingencies align with the brain’s expectations, leading users to experience virtual bodies as their own \citep{slater2009place}. The resulting sense of presence, the feeling of “being there”, can be understood as successful perceptual inference that the virtual world is sufficiently real to act within \citep{slater2022separate, triberti2025being}. Importantly, compelling experience depends not only on rendering fidelity but also on behavioural and narrative coherence, because incoherent cues can collapse plausibility even in highly immersive systems \citep{skarbez2020immersion}. This implies a transparency obligation that goes beyond “what was rendered”: systems should help users distinguish evidential cues from narrative framing, and support stepping out of persuasive framing when desired.

Empirical studies show that controlled manipulation of perceptual evidence can yield lasting changes in internal models. Rastelli et al. \citep{Rastelli2022VRhallucinations} found that VR-induced visual hallucinations increased cognitive flexibility by loosening entrenched perceptual priors, while \citep{Job2025VRexpectations} showed that a placebo VR intervention that visually exaggerated physical capability produced sustained real-world motor gains by elevating expectations about bodily ability. In predictive-processing terms, both effects reflect prior updating through structured prediction errors that can persist beyond the session. This strengthens the case for consentful, goal-aligned interventions that build competence rather than dependency, especially when altered expectations may generalise to everyday life.

Taken together, these examples show XR acting as both a perceptual enhancer and filter, enabling direct intervention in the inferential processes that generate perception by adding, removing, or restructuring sensory evidence. Such mediation parallels operant conditioning, where stimulus presence or removal guides learning and behaviour \citep{skinner2019behavior}, a mechanism already leveraged in XR design to intentionally engage or disengage users \citep{piumsomboon2022ex}. To avoid drifting from support into behavioural control, reinforcement intent should be explicit, auditable and user-configurable in high-stakes contexts.

XR thereby makes tangible the dependent-origination insight that experience is conditioned, while predictive processing and FEP explain how altered evidence reshapes inference over time \citep{ho2022intersubjectivity, parr2022active}. Recent AR work also operationalises SDT directly by testing how adaptive assistance shifts perceived autonomy: in AR-assisted construction assembly, low-agency control reduced workload but also reduced perceived autonomy, highlighting the agency trade-off that Self++ is designed to manage \citep{yang2025control}. Co-determination then specifies the interactional obligation: because XR can modify the conditions of experience, perceptual regulation should remain \textit{transparent} about what is changed and with what uncertainty, \textit{adaptive} as competence and context evolve, and \textit{negotiable} through consent, override, and reversion, in line with the co-determination principles (T.A.N.), so that uncertainty regulation supports autonomy, competence, and relatedness rather than bypassing them.

\subsection{Human--AI Interaction and Teaming in XR}

The preceding sections framed XR as a mechanism for intervening on the evidential conditions of experience (through perceptual filtering), and situated Self++ within a broader philosophical view in which experience and identity are enacted through relational conditions. When we move from perception to action, the locus of risk and opportunity shifts: the AI is no longer merely shaping what is seen or attended to, but increasingly participates in goal selection, planning, and execution. This transition places Self++ within Human--Agent Teaming (HAT) or Human--AI Teams (HATs), which studies how humans and autonomous systems coordinate to achieve shared goals \citep{licklider2008man, seeber2020machines, Zhang2021IdealHuman}. In XR and metaverse-like settings, teaming is not only informational but embodied and situated: coordination unfolds through shared spatial context, sensorimotor coupling, and the ongoing regulation of cognitive load and uncertainty.

A central obstacle for effective teaming is the user’s difficulty in forming accurate mental models of an agent’s state and reasoning, a challenge Norman characterises as the “gulf of evaluation” \citep{norman1988psychology}. In XR, this gulf can widen: immersive presentation may increase perceived immediacy and credibility, while the agent’s internal uncertainty, constraints, and operating assumptions remain hidden. This is precisely where the interactional stance of co-determination becomes necessary. Rather than assuming fixed tool use or unilateral automation, co-determination treats human and agent as joint participants in a coupled system, requiring that the agent’s intent, boundaries, and uncertainty be legible enough for the user to retain volitional control. Users also carry expectations about what a “good” AI teammate should be. \citep{Zhang2021IdealHuman} shows that people often expect AI partners to behave with human-like reliability, cooperativeness, and contextual sensitivity, and mismatches between these expectations and actual system behaviour can undermine trust and coordination. These expectation dynamics strengthen the case for co-determination as a stabilising baseline: the system must help users calibrate what the agent can and cannot do, rather than letting anthropomorphic assumptions silently drive reliance. Building on this trajectory, cognitive externalisation is now evolving into adaptive agent teammates, where calibrated trust shapes effective interaction and HAT outcomes \citep{duan2025trusting}.

In this practical domain of teaming, the co-determination principles (T.A.N.) must be implemented as specific interaction mechanisms rather than treated as abstract ethical principles:
\begin{itemize}
    \item \textbf{Transparency: bridging the gulf of evaluation.} The agent should make its internal state legible enough for users to form accurate mental models, including what it is optimising for, what it believes, and where uncertainty or constraints apply. This reduces evaluation gaps and supports trust calibration \citep{norman1988psychology, yousefi2025team, zhang2025hat}.
    \item \textbf{Adaptivity: dynamic allocation of initiative.} Effective teammates do not behave identically regardless of context. Agents should adjust initiative, timing, and level of autonomy as user confidence, workload, and task conditions change, supporting decision outcomes without overwhelming or bypassing the user \citep{Dong2024XRDecision}.
    \item \textbf{Negotiability: consensual delegation and recovery.} As agents become more capable, the risk of automation bias and loss of control increases. Users should be able to consent to actions, revise autonomy levels (e.g., ``help me do this'' versus ``do this for me''), and override or undo decisions, preserving authorship and accountability \citep{Han2026ExploringMediation}.
\end{itemize}

General human--AI interaction guidance reinforces these requirements. Established guidelines emphasise making clear why the system acted, supporting efficient correction, and enabling undo and refinement \citep{Amershi2019Guidelines}. In XR, where the system can shape both evidence and action, such principles are not cosmetic: they protect autonomy and competence by reducing surprise, supporting trust calibration, and preventing opaque shifts in control. Recent empirical work on team dynamics in human--AI collaboration further emphasises that teaming outcomes depend on interaction quality, affecting confidence, satisfaction, and accountability \citep{yousefi2025team}. From a Self++ perspective, these are not merely usability metrics; they indicate whether an agent supports or frustrates SDT needs, and whether the coupled system converges towards stable, low-surprise coordination.

This interactional framing is consistent with XR-specific work on explanation and intelligibility. The XAIR framework \citep{xu2023xair} for explainable AI in AR argues that systems should generate explanations with AI outcomes and keep them accessible to support user agency, while using manual, user-triggered delivery as the default due to limited cognitive capacity in AR. XAIR further recommends that automatic, just-in-time explanations be reserved for constrained cases (e.g., surprise or confusion, unfamiliar outcomes, or model uncertainty) and only when the user has enough capacity to attend to them. Beyond timing, XAIR emphasises end-user configuration and a longer-term user-in-the-loop co-learning process, where systems adapt to users while users’ understanding and AI literacy evolve. In HAT terms, these design commitments instantiate the co-determination principles (T.A.N.): explanation access and state-legibility as \textit{Transparency}, timing and initiative control as \textit{Adaptivity}, and user-trigger, configuration, and reversibility as \textit{Negotiability}.

Recent XR-specific HAT work further illustrates how embodied context changes the nature of coordination. Zhang et al.’s ``Virtual Triplets'' framework \citep{Zhang2024Triplets} analyses dynamics between the human, the virtual agent, and the physical task across synchronous and asynchronous settings. Successful assistance requires sensitivity to physical constraints, task progress, and translation between digital instruction and physical execution, aligning with the \textit{Competence} overlay of Self++: the agent’s role is not to replace skill, but to scaffold effective action \citep{vygotsky1978mind}. XR training research demonstrates this scaffolding role in practice. HAT Swapping \citep{zhang2025hat} explores how virtual agents can act as stand-ins for absent human instructors, enabling guidance and feedback to persist across time and personnel while preserving the structure of collaborative training. AVAGENT \citep{nam2025avagent} similarly shows how AI-powered virtual avatars can bridge asynchronous communication by capturing, transforming, and re-presenting human intent and context across time, extending HAT beyond real-time co-presence into persistent coordination in XR. Together, these systems highlight both the promise and responsibility of XR agents: they can reduce uncertainty and support skill acquisition, but only if guidance remains transparent, appropriately timed, and adjustable to the learner’s evolving competence.

As agents become more capable, the design challenge intensifies. Multimodal foundation models enable systems that can perceive and act across vision, audio, language, and contextual signals, supporting increasingly high-level delegation \citep{yang2025magma}. However, increased capability increases the risk of misalignment and opacity, especially when the user cannot inspect the agent’s evolving beliefs or intentions. Work on transparency for modern AI systems emphasises interactive scrutability, user education, and attention to socio-cultural context as prerequisites for trustworthy deployment \citep{liao2023ai}. In Self++ terms, this is a direct extension of co-determination: an agent that can act must remain accountable through transparency and negotiability, allowing users to query intent, adjust autonomy, and recover from errors. Approaches that explicitly structure agent reasoning, such as BDI-style models \citep{li2025satori}, can support this by making beliefs, desires, and intentions more inspectable, thereby narrowing the gulf of evaluation and reducing the likelihood that delegation undermines user agency.

These issues are not unique to XR, and lessons from human--AI co-creation generalise. Studies of collaborative writing with language models highlight recurring problems of trust calibration, user control, and authorship, even in ostensibly low-stakes tasks \citep{lee2022coauthor}. Recent work on agency in LLM-infused tools similarly suggests that preserving authorship depends on making suggestions legible and easy to veto, so that assistance remains subordinate to the user’s intent rather than silently steering outcomes \citep{nishal2025designing}. These findings map naturally onto the \textit{Autonomy} overlay of Self++ and provide concrete interaction criteria for co-determination in XR: the agent should behave as a co-pilot whose contributions are inspectable, revisable, and aligned with the user’s chosen level of control. Shneiderman’s Human--Centred AI perspective reinforces this stance, arguing for systems that combine powerful automation with meaningful human control and accountability rather than pursuing full autonomy as an end in itself \citep{shneiderman2020human}. \citep{yang2020re} likewise notes that human--AI interaction is uniquely difficult to design due to unpredictability, feedback gaps, and mental-model mismatch, highlighting the need for iterative design processes to tame agent behaviour in practice. In XR, where the system can shape both action and perception, these difficulties are amplified, strengthening the case for co-determined interaction as a baseline expectation.

Finally, the social dimension of teaming is essential, particularly for the \textit{Relatedness} overlay of Self++. Triadic human-agent dynamics show that agents can mediate human-human collaboration, influencing how people coordinate and communicate with one another \citep{Han2026ExploringMediation}. Embodied virtual agents can elicit prosocial responses by expressing social cues and emotional responsiveness, suggesting a pathway for agents to scaffold social functioning rather than merely simulate companionship \citep{yousefi2024advancing}. More generally, empirical work in AR suggests that visual embodiment and social behaviours shape how intelligent assistants are perceived, influencing credibility, social presence, and users’ willingness to engage with or rely on an agent \citep{kim2018does}. At the same time, work on ``generative agents'' shows how AI characters with memories and routines can produce richly believable social dynamics in simulated worlds \citep{park2023generative}. Such agents could enrich relatedness in XR by enabling meaningful social rehearsal, community participation, or culturally situated narratives; yet they also sharpen the risks of uncanny, unpredictable, or normatively misaligned behaviour that can destabilise user trust and undermine a sense of agency. From the philosophical framing established earlier, this is not peripheral: if experience and identity are enacted through relational conditions, then agent-mediated social worlds will shape the kinds of selves that become habitual. Co-determination, therefore, extends beyond individual task control to relational accountability: agents that participate in social coordination should be designed to strengthen users’ connection to real communities and shared norms, rather than displacing human relationships through frictionless substitutes \citep{liu2025heterogeneous}.

Taken together, HAT in XR offers the interactional mechanisms through which Self++ can be realised across the three overlays: competence, autonomy, and relatedness. XR can reorganise sensory evidence and reduce uncertainty, but as AI shifts from filter to collaborator, the conditions for healthy regulation of uncertainty become fundamentally interactional. Co-determination provides the bridge from the cognitive and philosophical foundations to concrete HAT practice: by prioritising the co-determination principles (T.A.N.), XR agents can scaffold skill, preserve volitional control, and strengthen social embeddedness, rather than causing relational displacement.

\section{The Self++ Architecture: Three Overlays of Augmented Agency}

Self++ organises human--AI coupling into three concurrently activatable \textit{overlays} (Self, Self+, Self++), forming an architecture of augmented agency. Each overlay targets a different temporal and functional scale of free-energy minimisation, corresponding to nested timescales of adaptation and echoing ``nested learning'' in AI \citep{behrouz2025nested}. The naming (Self, Self+, Self++) does not imply separate selves, but an expanding \emph{scope} of agency support: from here-and-now action, to deliberation and policy formation, to social embeddedness.

Importantly, Self++ does \emph{not} assume a strict pipeline in which Overlay 1 must finish before Overlay 2 or Overlay 3 begins. In realistic settings (training, teamwork, community participation), competence-building, autonomy exercise, and relatedness-support often co-occur, and initiative shifts fluidly between human and system \citep{Horvitz1999MixedInitiative, Bradshaw2003AdjustableAutonomy}. We therefore treat the three overlays as concurrently activatable modes of a coupled human--AI system, with different emphases depending on context and risk.

\begin{enumerate}
    \item \textbf{Overlay 1 (Self): Competence at the sensorimotor timescale (seconds to minutes).}
    This overlay augments perception and skill, reducing immediate prediction errors in action execution \citep{clark2013whatever}.\\
    \textit{Mechanistic coupling (SDT--FEP):} \textit{Competence $\longleftrightarrow$ minimisation of sensorimotor prediction error.}
    Competence is the subjective experience of a high-precision internal model effectively governing action. When the AI scaffolds skill (e.g., highlighting a target), it reduces the gap between predicted and actual sensory feedback, validating the user’s model of agency \citep{vygotsky1978mind, greunke2016taking}.

    \item \textbf{Overlay 2 (Self+): Autonomy at the deliberative and situational timescale (minutes to days).}
    This overlay augments cognition and decision-making, helping users navigate complex choices and intermediate goals by reducing strategic uncertainty \citep{Ryan2000SDT}.\\
    \textit{Mechanistic coupling (SDT--FEP):} \textit{Autonomy $\longleftrightarrow$ preservation of high-level priors (policy selection).}
    Autonomy reflects the ability to self-endorse actions. In FEP terms, this equates to selecting policies aligned with deep, top-down priors (values/goals), rather than being driven by bottom-up salience or external coercion \citep{Ryan2000SDT}. AI support here reduces ``decision entropy'' while protecting the user’s generative model from being overwritten by the system \citep{shneiderman2020human}.

    \item \textbf{Overlay 3 (Self++): Relatedness at the developmental and existential timescale (months to years).}
    This overlay augments social connection and purpose, steering long-term trajectories and relationships by aligning actions with enduring values and shared social models \citep{orlosky2021telelife}.\\
    \textit{Mechanistic coupling (SDT--FEP):} \textit{Relatedness $\longleftrightarrow$ alignment of shared generative models.}
    Relatedness arises from synchronisation of internal models between agents: social connection enables partial offloading of uncertainty onto the group. AI support here minimises ``social surprise'' (misinterpretation of others) and helps the user remain embedded in a shared communicative web \citep{jing2021eye, ho2022intersubjectivity}.
\end{enumerate}

A methodological note on these couplings: SDT and FEP operate at different levels of description; SDT is a motivational theory grounded in decades of experimental psychology, while FEP is a formal account of biological self-organisation rooted in variational inference. The correspondences proposed above (competence $\longleftrightarrow$ sensorimotor prediction-error minimisation; autonomy $\longleftrightarrow$ high-level prior preservation; relatedness $\longleftrightarrow$ shared generative-model alignment) are bridging hypotheses that constitute part of this paper's theoretical contribution, not established equivalences. We propose them because they generate productive design commitments (the overlays and T.A.N. constraints) and because they yield testable predictions (Table \ref{tab:selfpp_propositions}, \textit{P2} and \textit{P6} in particular). However, the mapping is neither unique nor exhaustive: alternative bridging constructs are possible, and empirical work may reveal that the coupling is tighter for some needs than others or that additional mediating constructs are required. We therefore treat these correspondences as working hypotheses to be refined through the evaluation programme outlined in Section \ref{sec:selfpp_propositions} and the future work discussed in Section \ref{sec:limitations-future-work}. 

Within each overlay, Self++ specifies three \textit{role patterns} (R1--R9 in total), each realised as an AI role that supports the user under co-determination. R1--R3 correspond to \textit{Tutor, Skill Builder, Coach} (Overlay 1); R4--R6 to \textit{Choice Architect, Advisor, Agentic Worker} (Overlay 2); and R7--R9 to \textit{Contextual Interpreter, Social Facilitator, Purpose Amplifier} (Overlay 3). Together, these nine role patterns form the core of the Self++ framework (summarised in Table \ref{tab:selfpp_roles}). The decomposition into named role patterns, rather than a single continuously adaptive agent or an undifferentiated spectrum of support, serves three design functions aligned with co-determination. First, discrete roles make the system's current supportive intent legible to the user: knowing that the system is acting as a Coach (testing robustness) rather than a Tutor (guiding a novice) allows the user to calibrate expectations and interpret feedback correctly, directly supporting Transparency. Second, bounded role definitions give designers and evaluators discrete interaction contracts against which to assess T.A.N. compliance and test the propositions in Table \ref{tab:selfpp_propositions}; a diffuse, unlabelled adaptation would be harder to audit or compare across implementations. Third, named roles provide natural anchors for Negotiability: users can request transitions ("stop advising, just execute"), decline specific modes ("no stress tests today"), or query the system's current stance ("why are you coaching rather than building?"), interactions that would be less intuitive if support varied along an unmarked continuum. We note that this decomposition is functional, not architectural: whether the nine roles are implemented as a single foundation model selecting behavioural policies, as composable modules, or as a mixture of both is an engineering decision left to implementation. What matters for Self++ is that the user-facing interaction contract of each role remains distinct and legible regardless of the underlying system design. While functional, roles are personified (e.g., Tutor, Coach, Teammate, Guide) to clarify supportive intent rather than authority.

Crucially, role patterns act as \textit{adaptive scaffolds}: as competence, context, and risk change, the system transitions between role patterns or fades support to prevent over-reliance and to preserve human autonomy and relationships \citep{turkle2011alone}. To keep augmentation legitimate rather than covert control, Self++ applies the \textbf{co-determination principles (T.A.N.)} across all overlays:

\begin{itemize}
    \item \textbf{Transparency:} Sufficient information for accurate mental models of intent, limits, incentives, and uncertainty.
    \item \textbf{Adaptivity:} Support tuned over time as competence and context evolve (including fading).
    \item \textbf{Negotiability:} Volition preserved via consent, override, and adjustable autonomy.
\end{itemize}

T.A.N. requirements strengthen with scope and initiative: higher-overlay role patterns (especially those touching identity, relationships, or long-horizon behaviour) demand stronger transparency and negotiability as safeguards \citep{Horvitz1999MixedInitiative, Bradshaw2003AdjustableAutonomy}.

\section{Overlay 1 – Foundational Augmentation of the Self (Competence Support)}

Overlay 1 targets competence at the sensorimotor timescale: helping users perceive and act reliably in an enriched environment, while keeping early errors and overload low enough for learning to take hold. Self++ does not treat this as a prerequisite pipeline stage. Competence support often runs in parallel with autonomy and relatedness supports (for example, training in teams), but Overlay 1 remains the point where the system most directly shapes perceptual evidence and action feedback.

Mechanistically, Overlay 1 reduces sensorimotor prediction error so users experience effectance and learnable control: attention is guided, actions are constrained into safe steps, and feedback tightens the link between intention and outcome. In SDT terms, this sustains competence by enabling early, attributable successes; in FEP terms, it increases the precision of action-outcome mappings and reduces surprise during control \citep{Ryan2000SDT, Friston2010FreeEnergy}. We define three role patterns that mirror established progressions in skill acquisition from novice to proficient performance: \textit{Tutor (R1)}, \textit{Skill Builder (R2)}, and \textit{Coach (R3)} \citep{dreyfus2004five}. As with higher overlays, these role patterns are co-determined and scaffolded in line with the co-determination principles (T.A.N.), where support should be \textit{Transparent} (users can tell what is being guided and why), \textit{Adaptive} (fading as competence stabilises), and \textit{Negotiable} (users can slow, pause, or override guidance), so assistance accelerates learning without converting into dependency.

\subsection{Role Pattern R1: Guided Familiarisation (AI as Tutor)}

At the outset of a new task or environment, novices face high uncertainty because relevant cues, action boundaries, and error consequences are not yet well-modelled. In the Tutor role, the AI adopts a proactive stance that structures the experience into a learnable corridor: it highlights what matters, suppresses what is distracting, and sequences actions so that each step is achievable before the next is introduced. This is classic scaffolding in the Zone of Proximal Development \citep{vygotsky1978mind}, but implemented through in-situ perceptual guidance rather than detached instructions.

In XR, this guidance can be spatial and embodied: key objects or regions can be highlighted, next actions can be indicated with anchored arrows \citep{lee2018live360} or ghosted exemplars \citep{oda2015replicas}, and irrelevant elements can be visually deemphasised to reduce split attention. A practical pattern is \emph{step gating}: the system reveals only the next required sub-action and advances when completion is detected, which keeps working memory demands bounded. Adaptive AR tutoring systems have operationalised this idea by monitoring tutorial-following status and adjusting the amount and form of guidance in real time \citep{huang2021adaptutar}. When attention lapses, a Tutor can also regulate pacing through attention-aware playback (for example, pausing or slowing guidance when gaze or location cues indicate the user has fallen out of sync), helping the user recover without compounding errors.

Technically, the Tutor role overlaps with intelligent tutoring systems that use cognitive models to interpret learner actions and deliver context-sensitive feedback (for example, model tracing and related methods in cognitive tutors) \citep{anderson1995cognitive}. The key difference in XR is that feedback can be embedded directly into the perceptual field, allowing guidance to be shown where and when it is needed rather than translated into verbal rules.

Empirical evidence supports the value of structured, in-situ guidance during early skill acquisition. In assembly-like tasks, AR instructions have been shown to reduce errors and improve performance relative to conventional instruction formats in controlled comparisons \citep{vanneste2020cognitive}. At the same time, the broader literature cautions that AR can either reduce or increase cognitive load depending on design choices, which strengthens the case for tightly scoped, well-timed guidance at R1 \citep{buchner2022impact}.

Finally, the Tutor role pattern must not become a permanent crutch. Self++, therefore, treats R1 as deliberately temporary: as soon as the user demonstrates stable performance on a step, guidance should begin to fade (fewer cues, larger action windows, more self-explanation) and the system should transition towards the Skill Builder role pattern. This aligns with evidence from instructional design that gradually reducing worked guidance supports the learner’s shift from example-following to independent problem solving \citep{atkinson2000studying, sweller2011guidance}. In summary, R1 establishes a safe learning corridor with immediate, situated success signals, while explicitly preparing the conditions for the scaffold to be removed.

\subsection{Role Pattern R2: Scaffolded Practice (AI as Skill Builder)}

Once the user can complete the basic sequence under guided familiarisation, the AI shifts into the Skill Builder role pattern that prioritises practice, calibration, and generalisation. The support envelope deliberately widens: the system provides partial cues and performance feedback, but stops prescribing every micro-action. The intent is to refine the user’s sensorimotor predictions while avoiding the brittleness that comes from rehearsing a single, fixed script. Motor-learning theory predicts that variability and appropriately structured interference during practice can improve transfer and retention, even if acquisition feels harder \citep{schmidt1975schema, Raviv2022variability}.

A hallmark of R2 is augmented feedback that keeps ``what good looks like'' visible while leaving execution to the user. Two common XR patterns are Ghost Tracks, which overlay time-aligned expert motion for in-situ trajectory and timing matching \citep{Thanyadit2022XRLive, dominguez2023dataset, cho2025Bichronous,yang2002justfollowme, jarc2017proctors}, and Shadow Workspaces, which anchor a target end-state silhouette (``shadow of success'') to support precise pose, placement, or orientation \citep{Piumsomboon2014Grasp, Thanyadit2022XRLive, Limbu2018sensors, buchner2022impact}. Together, they externalise comparison and reduce cognitive load during repeated practice while preserving active control.

Although these cues are most natural for 3D sensorimotor tasks, the underlying principle generalises: externalised reference structure reduces internal memory and computation by making intermediate steps, trajectories, or goal states inspectable \citep{kirsh2010thinking}. In non-spatial skills, this parallels worked examples and step-by-step solution traces that support novice learning and later independent problem solving \citep{kirschner2018cognitive, renkl2014worked}. Seen this way, Ghost and Shadow patterns are XR instantiations of a broader apprenticeship logic of modelling, scaffolding, and fading across embodied and cognitive procedures \citep{collins2018cognitive, anderson1995cognitive}.

Crucially, R2 also introduces controlled challenge. Rather than maximising ease, the system should keep the task in a learnable difficulty band by gradually withholding hints, expanding acceptable action ranges, and introducing mild perturbations (for example, small changes in order, timing constraints, or plausible micro-faults) so the user learns to adapt rather than imitate. This ``challenge just beyond current mastery'' is consistent with the challenge-skill balance emphasised in flow-oriented accounts of engagement and growth \citep{csikszentmihalyi1990flow}. It also parallels curriculum ideas from machine learning, where a teacher proposes goals that are increasingly difficult but achievable, as in AMIGo \citep{campero2021amigo}. In Self++, the Skill Builder role pattern therefore balances error reduction with productive difficulty: enough structure to prevent unproductive surprise, enough freedom and variability to build robust competence. By the end of R2, the user should rely on Ghost and Shadow cues primarily for fine-tuning, while completing substantial portions of the task without explicit step-by-step prompting.

\subsection{Role Pattern R3: Mastery and Resilience (AI as Coach)} 
\label{sec:r3-coach}

Once the user is reliably proficient in routine conditions, the AI transitions to the Coach role pattern focused on robustness, adaptability, and self-correction. Guidance recedes: instead of persistent highlights or continuous overlays, the Coach monitors performance and introduces controlled perturbations to test whether the skill generalises beyond rehearsed cases. This deliberate use of ``desirable difficulties'' supports more durable, flexible learning than perfectly predictable practice \citep{bjork2011making, landman2018training} and matches accounts of expertise that emphasise deliberate, feedback-rich refinement over time \citep{ericsson1993role}.

In practice, the Coach varies scenarios, injects plausible faults, and occasionally withholds support (for example, removing an overlay or altering timing constraints) to expose brittle assumptions and reveal blind spots. It intervenes only when performance drops below a safety or quality threshold, preventing the consolidation of poor habits while keeping the user responsible for recovery and strategy. After each episode, the Coach provides a brief debrief and, if needed, temporarily reverts to Tutor or Skill Builder to remediate a specific sub-skill. In Self++ terms, R3 consolidates competence by reducing ``surprise under stress'': the user learns not only to execute correctly, but to remain stable when conditions deviate from expectation \citep{landman2018training}.

R3 also manages role transitions in team settings, so the user retains a coherent model of who is doing what. Abrupt hand-offs, silent autonomy shifts, or ambiguous identities can trigger mode confusion and automation surprise, especially in off-nominal situations \citep{sarter1995mode, eom2022modeconfusion}. Self++ operationalises this as HAT Swapping \citep{zhang2025hat}: a user-legible protocol for transferring a functional role between human and AI teammates (and back again), preserving continuity cues where helpful while explicitly disclosing changes in agency and identity so trust remains calibrated \citep{lyons2021humanautonomy}. When the coach function is embodied (for example, via an avatar), continuity cues (voice, interaction style, interface layout) can smooth transitions, but disclosure of boundaries and identity remains essential, in line with the co-determination principles (T.A.N.): \textit{Transparency} about what changed and why; \textit{Adaptivity} in how much coaching is offered; and \textit{Negotiability} via consent for stress tests, override, and adjustable autonomy. Trust calibration work suggests these disclosures and capability cues help prevent over- or under-reliance \citep{wischnewski2023trust, okamura2020trustcalibration}.

By the end of R3, the user should display functional mastery: resilient performance across varied conditions, recovery from errors without constant prompting, and correctly calibrated trust in the coach as a safety net rather than a crutch.

\section{Overlay 2 – Cognitive and Strategic Augmentation (Autonomy Support)}

Overlay 2 shifts emphasis from executing skills to forming and revising policies: choosing goals, weighing trade-offs, and allocating attention and effort over time. Self++ does not treat the three overlays as a strict pipeline. Autonomy support often appears \emph{during} competence building: even in training, learners must make meaningful choices (what to try next, when to speed up, whether to accept risk, when to request help) in order to demonstrate genuine competence. Accordingly, Overlay 2 can run concurrently with Overlay 1: the system may coach sensorimotor execution while also shaping the user’s decision context so choices remain aligned with the user’s own values and intentions.

This concurrent view matches mixed-initiative and adjustable-autonomy systems, where initiative and control shift fluidly between human and agent depending on task demands, user state, and risk, rather than advancing through fixed stages \citep{Horvitz1999MixedInitiative, Bradshaw2003AdjustableAutonomy, Beer2014AutonomyFramework}. Mechanistically, Overlay 2 targets autonomy as policy selection: in SDT, autonomy is experienced as self-endorsed action \citep{Ryan2000SDT}; in FEP terms, this corresponds to protecting high-level priors (values and goals) while using prediction to reduce uncertainty about consequences \citep{Friston2010FreeEnergy}. In Self++ terms, Overlay 2 is co-determination expressed at the cognitive timescale: a second voice that helps the user reflect, anticipate outcomes, and surface trade-offs, but does not smuggle in new goals or override the user’s higher-order commitments. This caution is reinforced by evidence that synthetic persuasion evaluations can diverge from human outcomes \citep{doudkin2025aipersuading, doudkin2025synthetichumangap}.

A key autonomy risk in modern ecosystems is that choice environments are routinely shaped by opaque recommendation logic, engagement optimisation, and dark-pattern design, steering behaviour while eroding the user’s sense of authorship \citep{mathur2019darkpatterns, luguri2021darkpatterns}. Self++, therefore, requires decision support to remain co-determined: (i) legible enough for users to judge how the system is weighing attention and effort, (ii) responsive to changing goals and context, and (iii) subject to consent, override, and adjustable autonomy. This reflects long-standing guidance that automation should act as a collaborative partner rather than an invisible controller \citep{klein2004teamplayer, shneiderman2020human}.

We define three role patterns in this overlay as \textit{Choice Architect (R4)}, \textit{Advisor (R5)}, and \textit{Agentic Worker (R6)}, reflecting increasing initiative in shaping the decision environment, explaining trade-offs, and executing actions, but always under user oversight, reversibility, and the co-determination principles (T.A.N.) introduced earlier.

\subsection{Role Pattern R4: Subtle Guidance in Choice (AI as Choice Architect)}

At R4, the AI begins to shape the \textit{decision context} rather than the decision itself. As a \textit{Choice Architect}, it uses small changes in salience and friction to make goal-consistent options easier to notice and compare while leaving selection entirely with the user. This draws on classic choice architecture and nudging, but under a stricter co-determination constraint: the system may guide attention, but must not covertly redirect goals or exploit vulnerabilities \citep{thaler2008nudge, sunstein2015nudging, schmidt2020ethics}. In XR, this can be enacted through lightweight perceptual cueing, for example, gently highlighting items that match the user’s stated dietary goal in an AR aisle \citep{Dong2024XRDecision}, or rendering a user-preferred route as more visually salient via in-view AR guidance while leaving all alternatives selectable \citep{tonnis2008perceptionthresholds, kim2009windshield}.

Mechanistically, this role pattern operates by re-weighting attentional evidence: the interface makes some cues more precise (more noticeable, easier to act on) so that acting on existing intentions requires less search and self-control. Because the same mechanism can become manipulation, R4 should be treated as scaffolding for autonomy, not behaviour steering. Self++ therefore binds Choice Architect nudges to co-determination principles (T.A.N.) safeguards: \textit{Transparency} that the highlight is system-generated and why, \textit{Adaptivity} that tracks the user’s changing priorities rather than a single platform metric, and \textit{Negotiability} through opt-out, adjustable strength, and consent for high-stakes nudges \citep{sunstein2015nudging, meske2020ethicalnudges}. These safeguards are especially important when R4 is running concurrently with Overlay 1 coaching, because the learner’s heightened reliance and reduced situational bandwidth can otherwise make helpful layout indistinguishable from hidden coercion.

Finally, implementing Choice Architect support requires multi-objective reasoning: most real decisions trade off plural values (cost, safety, enjoyment, time), so the system should represent trade-offs and let the user steer weights rather than collapsing everything into an opaque score \citep{sorensen2024pluralisticalignment}. In this way, R4 reduces decision friction and strategic uncertainty while preserving experienced authorship: the user can always recognise, contest, and revise how the system is shaping the field of choice.

This design stance also clarifies the ethical status of nudging within R4. The nudge debate has shown that whether a nudge is manipulative depends less on the inevitability of framing decisions and more on the mechanisms by which the nudging occurs and whether the direction of influence is transparent to the target \citep{sunstein2015nudging, schmidt2020ethics, noggle2022manipulation}. Self++ resolves this tension procedurally: every nudge in R4 must be transparently marked as system-generated and linked to the user's own stated goals, adaptively tuned to changing priorities rather than a fixed platform metric, and negotiable through opt-out, adjustable strength, and consent gates for high-stakes choices (Table \ref{tab:selfpp_roles}). The system's influence therefore operates not by bypassing or subverting rational deliberation, but by scaffolding it—providing the informational and attentional conditions under which the user can deliberate more effectively while retaining full authorship over the resulting decision.

\subsection{Role Pattern R5: Informed Deliberation (AI as Advisor)} 

Where R4 shapes the \emph{choice environment}, R5 externalises the \emph{deliberation itself}. The AI becomes an Advisor: a conversational analyst that helps the user surface assumptions, compare futures, and reason through trade-offs, while keeping policy selection and endorsement with the user \citep{shneiderman2020human, reicherts2025ai}. This role pattern is especially important in contexts where persuasive optimisation can outperform genuine behaviour change in simulation but fail to translate into durable, owned decisions in the real world \citep{doudkin2025aipersuading, doudkin2025synthetichumangap}. In Self++, the Advisor is designed to feel like a co-determining voice that sharpens reflection rather than a persuader that steers outcomes.

Concretely, the Advisor provides \textit{interactive evidence and counterfactuals} rather than a single ``best'' answer. It can assemble an XR dashboard that contrasts options across the user’s stated criteria (for example, work-life balance, skill growth, risk, and social commitments), and allow the user to interrogate ``why'' and ``what if'' in place \citep{shneiderman2020human, haque2023xaiuserperspective}. A ``day in the life'' walkthrough, uncertainty bands, or side-by-side consequence traces can make long-horizon implications more legible without collapsing plural values into one score. The Advisor can also act as a memory and consistency check (``you previously prioritised family time''), and make second-order effects explicit (``skipping this meeting increases the chance of delaying Project X'') so the user is choosing with clearer foresight, not narrower freedom \citep{steyvers2023threechallenges, krakowski2025agency}.

R5, therefore, targets autonomy in its stronger sense: informed self-endorsement. It reduces ``decision entropy'' by illuminating unknowns and disagreements between objectives, but it must do so in line with the co-determination principles (T.A.N.). \textit{Transparency} requires surfacing data provenance, assumptions, and uncertainty (and what the model cannot know). \textit{Adaptivity} requires tuning explanation depth and modality to the user’s expertise and momentary cognitive load. \textit{Negotiability} requires editable goals, weights, and constraints, plus the ability to decline lines of reasoning, request alternatives, and override defaults. Together, these safeguards keep the Advisor supportive, legible, and revisable, so the user remains the author of the decision even when the AI is doing substantial analytic work \citep{shneiderman2020human, reicherts2025ai, Amershi2019Guidelines}.

 \subsection{Role Pattern R6: Empowered Delegation (AI as Agentic Worker)}

If R5 externalises deliberation, R6 externalises execution. Here, the AI becomes an \textit{Agentic Worker}: it carries out well-scoped tasks on the user’s behalf while remaining subordinate to the user’s intent and oversight \citep{krakowski2025agency, shneiderman2020human}. The user delegates an outcome (and constraints), the AI proposes an executable plan, and the pair iterates until the plan is endorsed. This preserves autonomy because the AI’s agency is not an independent authority, but an operational extension of the user’s chosen policy.

Because delegation increases the risk of out-of-the-loop failures, complacency, and automation surprise, R6 requires explicit safeguards \citep{endsley2017fromhere, endsley1995outoftheloop}. Precisely, the Agentic Worker should operate as a \textit{proposal-approval loop}: it presents what it intends to do (steps, assumptions, dependencies, and uncertainty), requests confirmation at appropriate checkpoints, and remains interruptible throughout \citep{shneiderman2020human, cheng2026safeagents}. Intermediate autonomy is preferred over set-and-forget automation: maintaining user involvement at key junctures supports situation awareness and improves recovery when the environment deviates from expectations \citep{endsley1995outoftheloop, kaber1997ootl}.

Self++ implements these safeguards through the co-determination principles (T.A.N.). \textit{Transparency} means the AI makes its intent, limits, and current authority legible (what it is doing, why, and what could go wrong). \textit{Adaptivity} means autonomy is adjustable and can be tightened or loosened as the user’s confidence, task criticality, and context change (for example, more confirmations for novel or high-stakes steps). \textit{Negotiability} means delegation is always explicit, revocable, and renegotiable: the user can override, pause, or re-scope the task at any time, and the AI treats corrections as first-class inputs rather than friction \citep{johnson2011coactive, shneiderman2020human}. This keeps the system aligned with the user’s values while reducing the need for persuasion; behaviour change is owned by the user because action follows endorsement, not covert steering \citep{doudkin2025aipersuading, doudkin2025synthetichumangap}.

At the end of R6, Overlay 2 reaches its apex: the user experiences \textit{augmented autonomy} in the strict sense; they remain the author of goals and approvals, while the AI reliably executes across tools and contexts with minimal cognitive burden \citep{krakowski2025agency, shneiderman2020human}. The result is higher throughput without surrendering control: autonomy is strengthened through delegation that is transparent, adjustable, and always negotiable \citep{endsley2017fromhere}.


\section{Overlay 3 – Societal and Existential Augmentation (Relatedness and Purpose)}
\label{sec:overlay3}

Overlay 3 moves into the most aspirational domain of augmented agency: supporting relatedness, cultural embeddedness, and long-horizon coherence with values and purpose. If Overlay 1 supports ``doing things right'' (competence) and Overlay 2 supports ``doing the right things'' (autonomy), Overlay 3 supports ``being the right self'' in relation to others: sustaining relationships, repairing misunderstandings, and maintaining value-aligned trajectories in social and civic life. Importantly, these supports are not strictly sequential; in real-world settings, like project teams or classrooms, competence-building and autonomy often unfold within the context of social coordination and conflict. Overlay 3 is therefore a high-scope overlay that stabilises shared meaning and belonging while Overlays 1 and 2 operate in parallel.

Mechanistically, Overlay 3 targets uncertainty at the level of shared generative models. Teams and communities function best when participants converge on shared mental models, a mutual understanding of ``what is going on'' and ``who is responsible for what'' \citep{mathieu2000influence, dedreu2003taskrelationship, dechurch2010cognitive}. At larger scales, collective sensemaking under information overload becomes a coordination problem where actors must distinguish signal from noise to avoid fragmentation \citep{weick1995sensemaking}. In SDT terms, this overlay protects relatedness by reducing social misattunement; in FEP terms, it minimises ``societal'' and ``existential'' entropy by ensuring that interacting agents remain aligned across both immediate actions and long-term timescales of meaning \citep{Ryan2000SDT, Friston2010FreeEnergy}.

Accordingly, Overlay 3 defines three role patterns mapping to R7--R9: \textit{Contextual Interpreter} (making identity, norms, and downstream impacts legible to prevent social surprise); \textit{Social Facilitator} (nurturing shared understanding and constructive conflict repair); \textit{Purpose Amplifier} (supporting value-aligned self-regulation and life coherence to prevent value drift) \citep{krakowski2025agency, sorensen2024pluralisticalignment}. Because these role patterns touch the core of identity, co-determination is non-negotiable. The system must act as a user-legible partner, not a hidden governor. We therefore apply the co-determination principles (T.A.N.) as a hard constraint, requiring explicit transparency and negotiability whenever the system intervenes in relationships, values, or civic judgment \citep{shneiderman2020human, doudkin2025aipersuading}.

 \subsection{Role Pattern R7: Big-Picture Contextualisation (AI as Contextual Interpreter)}

R7 addresses a recurring failure mode of hybrid XR--AI settings: people can act locally (and fluently) while lacking context about identities, roles, norms, provenance, and downstream consequences. The \textit{Contextual Interpreter} augments the user with situational and value-relevant legibility across two fronts: it surfaces information that may carry ethical, social, or practical significance for the user, without presupposing which normative framework applies. What counts as value-relevant is shaped by user configuration, cultural context, and the co-determination principles (T.A.N.), ensuring that context augmentation functions as epistemic support, expanding what the user can notice and anticipate, rather than as moral instruction.

On the social side, the Interpreter enforces \textit{identity and role clarity} in mixed human--AI ecologies. In XR meetings or co-learning scenarios, it should make agent identity and function legible (for example, persistent labels or outlines that distinguish humans from AI agents and indicate the current role, such as ``AI facilitator'' or ``human lead''). This is not cosmetic: disclosure cues help users calibrate expectations and preserve trust when agency shifts, including hand-offs structured via \textit{HAT Swapping}\citep{zhang2025hat}. Evidence from AI service contexts suggests that identity disclosure can measurably shape user trust and uptake, reinforcing the need for explicit signalling rather than ambiguity \citep{CHEN2025spotlight}. Beyond identity, the Interpreter can recover social signals that are weakened in mediated interaction (for example, shared gaze or attention cues), supporting mutual awareness and coordination \citep{jing2021eye, lukosch2015collaboration}.

On the world side, the Interpreter bridges micro-actions to macro-consequences without coercion. It can surface value-relevant context that would otherwise be invisible at the moment of choice (for example, lifecycle or stakeholder impacts, long-horizon trade-offs, or alignment with the user’s stated commitments), framed in the user’s own value language where appropriate. This is particularly relevant when users confront large-scale, contested topics (public sentiment, politics, international relations), where sensemaking is shaped by information overload and social dynamics. Here, the Interpreter should act as a \textit{plural-context} aid: exposing uncertainty, provenance, and credible alternative interpretations rather than optimising for a single persuasive outcome \citep{weick1995sensemaking, doudkin2025aipersuading}. In Self++ terms, the Interpreter reduces ``social surprise'' and ``contextual regret'': fewer breakdowns caused by misidentifying counterparts, misreading norms, or discovering too late that an action conflicted with one’s endorsed values.

Co-determination requirements are strongest in R7. \textit{Transparency} requires identity disclosure, provenance cues, and uncertainty communication; \textit{Adaptivity} requires tuning context density to attention and stakes (and backing off when low-value); and \textit{Negotiability} requires user control over what contexts are surfaced, when, and at what sensitivity, including opt-out and override. Together, these safeguards ensure that context augmentation functions as user-aligned sensemaking support rather than covert social steering \citep{shneiderman2020human, doudkin2025synthetichumangap}.

\subsection{Role Pattern R8: Facilitating Social Connection (AI as Social Facilitator)}

R8 moves from making context legible (R7) to actively improving how people relate and collaborate. Because real-world work, in classrooms, multidisciplinary teams, or cross-cultural communities, is inherently social, this role pattern often runs in parallel with competence-building (R1--R3) and autonomy support (R4--R6). Here, the AI acts not as a private companion, but as a light-touch facilitator that strengthens human-to-human coordination. By using XR to surface otherwise-missed social signals, it reduces the small misunderstandings that typically accumulate into conflict \citep{lukosch2015collaboration, orlosky2021telelife}.

The Social Facilitator builds shared mental models by restoring the attentional and intent signals often lost in mediated interaction. XR research demonstrates that cues such as gaze visualisation and mixed-reality communication markers can significantly improve grounding and social presence \citep{piumsomboon2018mini, piumsomboon2019effects, kim2019evaluating}. The Facilitator extends this by visualising group dynamics, such as participation balance or conversational rhythm, allowing teams to self-correct without a human moderator \citep{Han2026ExploringMediation, kim2008meetingmediator}. In fast-moving or jargon-heavy environments, the AI maintains common ground through optional micro-clarifications and role reminders, ensuring that shared understanding is actively supported rather than merely assumed \citep{yousefi2025team}.

Where friction arises, the AI defaults to process support, summarising viewpoints and prompting perspective-taking, rather than adjudicating outcomes. This focus on conversational flow is critical, as fast responsiveness is tightly linked to felt connection \citep{templeton2022fastresponse}. Crucially, Self++ treats R8 as explicitly pro-social: it aims to increase human-to-human contact rather than becoming the user’s primary relationship. While AI companions can reduce loneliness by making users feel ``heard'' \citep{defreitas2025aicompanions}, they also pose a risk of social drift toward synthetic companionship \citep{turkle2011alone}. The Social Facilitator mitigates this by preferentially scaffolding real-world relationships, inviting others in and encouraging repair after ruptures, and fading its own presence as human ties strengthen.

Because R8 touches group power and identity, it must be constrained by the co-determination principles (T.A.N.). \textbf{Transparency} requires clarity about what social signals are being sensed (e.g., ``is the AI tracking my tone?'') and how feedback cues are generated. \textit{Adaptivity} requires that the system can ``read the room'' and enter a do-nothing state when the group is thriving and intervention would be intrusive. \textit{Negotiability} must be socially contextualised, moving beyond individual consent to collective agreement. Specifically, \textit{Collective Negotiability} should offer: \textit{Mutual opt-in}, shared visualisations (like participation heatmaps) only appear if all members consent; \textit{Privacy-by-role}, allowing individuals to opt out of certain group metrics without social penalty; and \textit{Adjustable mediation}, enabling the group to negotiate the facilitation sensitivity, deciding, for example, whether the AI should flag interruptions or stay silent during heated creative debates. By situating negotiability within the group, the AI remains a tool for team co-determination, preventing failure modes like perceived surveillance or inadvertent shaming via data, and ensuring it remains a partner in human attunement.


\begin{table*}[t!]
\centering
\small 
\setlength{\tabcolsep}{2pt} 
\renewcommand{\arraystretch}{1.3} 

\caption{Self++ role patterns across overlays (concurrently activatable overlays) with example XR-AI behaviours and co-determination principles (T.A.N.).}
\label{tab:selfpp_roles}

    \resizebox{\textwidth}{!}{
        \begin{tabularx}{\textwidth}{ 
        @{}c l
        >{\hsize=1.0\hsize}Y
        >{\hsize=1.5\hsize}Y
        >{\hsize=0.8\hsize}Y
        >{\hsize=0.8\hsize}Y
        >{\hsize=0.8\hsize}Y@{}}
        \toprule
        \textbf{Lvl} & \textbf{Role} & \textbf{Role objective} & \textbf{Example XR-AI behaviours} & \textbf{Transparent} & \textbf{Adaptive} & \textbf{Negotiable} \\
        \midrule
        
        \multicolumn{7}{@{}l@{}}{\textbf{Overlay 1 (Self): Competence support (seconds to minutes)}} \\
        \midrule
        R1 & Tutor & Reduce novice uncertainty; establish safe learnable corridor & Anchored arrows and ghosted exemplars with step gating; clutter suppression; completion detection with attention-aware pacing and corrective feedback & Cue provenance; disclose suppression; show limits & Fade prompts; retarget errors; adjust pacing & Pause/skip; show all vs minimal; override highlights \\
        \addlinespace
        R2 & \makecell[l]{Skill \\ Builder}  & Calibrate + generalise; variability with feedback, not scripting & Ghost tracks and shadow end-states with partial hints; performance analytics with adaptive hinting and controlled variability & Explain feedback basis; show comparison model & Increase task variability; withhold hints; change modality & User-set difficulty; toggle ghosts; consent for perturbations \\
        \addlinespace
        R3 & Coach & Robustness under stress; self-correction; prevent brittle mastery & Fault injection and overlay removal with altered timing; safety/quality monitoring; targeted debrief with fall-back to R1/R2 & Disclose perturbation intent; disclose role/agency shifts & Adjust challenge intensity; adapt thresholds; taper monitoring & Opt-in for stress tests; emergency stop; hand-off confirmation \\
        \midrule
        
        \multicolumn{7}{@{}l@{}}{\textbf{Overlay 2 (Self+): Autonomy support (minutes to days)}} \\
        \midrule
        R4 & \makecell[l]{Choice \\ Architect}  & Shape decision context (salience) while preserving authorship & Lightweight cueing with route salience (alternatives remain selectable); multi-criteria filtering; attention reweighting with trade-off previews & Mark nudges; link to goals; label optimised criteria & Update weights; fade as user internalises; reduce during load & Opt-out slider; consent for high-stakes; unnudged view \\
        \addlinespace
        R5 & Advisor & Externalise deliberation; make counterfactuals inspectable & Interactive dashboards with side-by-side futures and uncertainty bands; value elicitation; model explanation with alternatives and effect highlights & Expose sources; distinguish evidence vs framing; show unknowns & Tune depth; switch modality; calibrate to time pressure & Editable goals; ask-for-alt; decline reasoning; override defaults \\
        \addlinespace
        R6 & \makecell[l]{Agentic \\ Worker} & Delegated execution under user policy; proposal-approval loop & Plan and execute with XR review checkpoints; plan trace with progress visibility; step confirmation with safe interrupts and rollback & Show intent/plan; audit trail; capability limits; risk disclosure & Adjust frequency by stakes; learn checkpoints; degrade gracefully & Explicit delegation; revoke anytime; re-scope; adjustable autonomy \\
        \midrule
        
        \multicolumn{7}{@{}l@{}}{\textbf{Overlay 3 (Self++): Relatedness \& purpose (months to years)}} \\
        \midrule
        R7 & \makecell[l]{Contextual \\ Interpreter}  & Legibility of identity/norms + impacts; reduce social surprise & Human vs AI labels and role badges; provenance overlays with impact cards; norm reminders; plural framing for contested topics & Radical disclosure of agent identity; show provenance & Context density tuned to attention; adapt to culture/values & Controls for context appearance; sensitivity sliders; opt-out \\
        \addlinespace
        R8 & \makecell[l]{Social \\ Facilitator}  & Improve coordination + repair; increase human-human connection & Shared gaze and participation balance visualisation; micro-clarifications; breakdown detection with viewpoint summaries and perspective-taking prompts & Disclose sensing granularity; explain prompts + thresholds & Do-nothing mode when thriving; calibrate to group norms & \makecell[l]{Collective opt-in;\\privacy-by-role;\\group-negotiable} \\
        \addlinespace
        R9 & \makecell[l]{Purpose \\ Amplifier}  & Long-horizon value coherence; steer away from disavowed futures & Value-facing simulations with nudges-in-narrative and framing controls; periodic reflections; contestable inferences with governance hooks & Reason + framing legibility; evidence vs narrative separation & Internalisation-focused fading; calibrate identity strength & Contestability; escalation requires opt-in; collective pathways \\
        \bottomrule
        \end{tabularx}
    }
\end{table*}

\subsection{Role Pattern R9: Aligning Life and Values (AI as Purpose Amplifier)}

R9 is the most delicate form of augmentation: the AI supports the user in living consistently with their self-endorsed values over long horizons, closing the gap between ``the life I intend'' and ``the life I drift into''. This targets long-timescale misalignment (chronic regret, value drift, attention capture) that can accumulate into ``existential surprise''. In SDT terms, the aim is not compliance but sustained autonomous self-regulation, where behaviour is owned and integrated rather than externally pressured \citep{Ryan2000SDT, Deci2012SDT}. In FEP terms, the \textit{Purpose Amplifier} helps the user maintain stable high-level priors (values and identity commitments) while flexibly updating mid-level plans and habits, reducing long-run expected free energy by steering away from trajectories the user later disavows \citep{Friston2010FreeEnergy, parr2022active}.

XR matters in R9 because immersive systems can intervene on the evidential stream that updates self and social priors, and therefore can reshape what the user comes to expect of themselves and others. Self-representation effects make this concrete: embodiment can shift attitudes and self-models in ways that generalise beyond the session (e.g., reductions in implicit bias following avatar embodiment) \citep{peck2013putting}. R9 interventions also often rely on \emph{nudges-in-narrative} (for example, story-consistent exit cues or value-aligned prompts embedded in a virtual routine). Here, coherence becomes an ethical boundary condition: if users cannot distinguish evidential cues from narrative framing, persuasion risks becoming covert control, even when intentions are benevolent \citep{skarbez2020immersion}; these risks sharpen in high-realism XR, motivating stronger safeguards for long-horizon behavioural shaping \citep{Slater2020EthicsVR}. At the same time, XR can make long-horizon consequences and value conflicts perceptible rather than abstract: whereas R8 strengthens relationships and group functioning in situ, R9 uses XR as a value-facing \emph{perceptual regulator} that externalises future selves, counterfactuals, and downstream impacts so the user can more reliably predict trade-offs and enact self-endorsed commitments \citep{pataranutaporn2024future}. For example, immersive encounters with age-progressed future selves can shift intertemporal choices towards long-term benefits \citep{hershfield2011saving}; VR perspective-taking can change social attitudes and prosocial tendencies by making another standpoint experientially salient \citep{Chen2023empathy}; and immersive climate experiences can improve learning and, in some settings, influence behavioural intentions and engagement by rendering invisible dynamics (e.g., ocean acidification) into lived evidence \citep{markowitz2018climatevr}. These are not prescriptions of ``what to value''; they are epistemic interventions that expand what the user can notice, anticipate, and contest, so value-consistent self-regulation becomes easier to sustain.

Practically, the system makes value-relevant discrepancies legible and actionable without turning them into coercion. It can surface periodic, user-configured reflections (e.g., how time, relationships, learning, and health track relative to stated priorities) and offer consentful, adjustable interventions aligned with the SPINED spectrum described earlier \citep{piumsomboon2022ex}. The default is the least forceful effective move: inform, nudge, or entice before deter, suppress, or punish. This matters because heavy-handed control risks undermining autonomy even when it improves short-term behaviour \citep{Ryan2000SDT}. The design centre of gravity remains co-determination: the AI functions as another voice in the user’s deliberative ecology, amplifying what the user has already endorsed, not substituting its own normative agenda \citep{shneiderman2020human}.

What changes in R9 is that co-determination expands beyond the individual human--AI dyad to the \emph{socio-technical loop} \citep{Herrmann2022socio} that shapes the dyad. R8 focuses on strengthening relationships and group functioning in situ; R9 governs the longer-run co-evolution of self, AI, and society by managing how systems shape preferences, norms, and incentives over time. Because XR interfaces can couple identity, attention, and affect into persuasive world-building, long-horizon alignment must treat recommendation and narrative loops as a coupled control problem: individual values guide system behaviour, system behaviour reshapes individual and collective priors, and institutions set reward structures that guide systems. This role makes such loops visible and steerable across two levels: personal settings for individual agency, and shared governance for teams and communities.

In R9, the co-determination principles (T.A.N.) become a \emph{societal} as well as personal constraint, and they must be sharper than in earlier roles because the intervention surface now includes identity cues, narrative framing, and institutional incentives:

\begin{itemize}
    \item \textbf{Transparency (reasons, framing, and incentives):} Interventions must include explicit ``because'' links to user-endorsed values (and visibility into what data is used, what is inferred, what incentives are optimised, and what uncertainty remains) \citep{shneiderman2020human, doudkin2025aipersuading, doudkin2025synthetichumangap}. In XR, transparency also requires \emph{framing legibility}: users should be able to inspect when a cue is narrative scaffolding versus evidential guidance, because coherence failures can become ethical failures \citep{skarbez2020immersion, Slater2020EthicsVR}.
    \item \textbf{Adaptivity (internalisation, not outsourcing):} The system should learn which supports feel autonomy-supportive (vs controlling), tune intensity and timing, and deliberately \emph{fade} scaffolds so the user internalises routines rather than outsourcing self-regulation indefinitely \citep{Ryan2000SDT}. In XR, adaptivity also means calibrating how strongly self-representation or narrative devices are used, since these can update priors about self and others \citep{peck2013putting}.
    \item \textbf{Negotiability (contestable boundaries and escalation control):} Override must extend from moment-to-moment control (``not now'', ``ask first'', ``reduce frequency'') to \emph{contestability}: users can challenge inferences, disable classes of interventions (e.g., identity-shaping cues, affective nudges), and demand alternative framings or evidence, with pathways for collective contestation when systems affect communities \citep{Leofante2024ContestableAI}. Any escalation in intrusiveness (towards deter/suppress) requires explicit opt-in and reversible settings \citep{piumsomboon2022ex}.
\end{itemize}

These safeguards are also stability conditions against metric pathologies. When proxies become targets, they invite distortion and strategic behaviour (by systems and by users), captured by Goodhart’s and Campbell’s laws \citep{ashton2021causal, Karwowski2023GoodhartRL}. R9, therefore, avoids single-score optimisation (e.g., ``screen time'' alone) as the governing objective. Instead, it treats wellbeing as plural, revisable, and context-dependent, keeping the user (and, when appropriate, the group) in the loop of redefining what counts as success \citep{sorensen2024pluralisticalignment, shneiderman2020human}. This closes the ethical loop: co-determination is not only a usability preference but a long-horizon alignment requirement.

Finally, R9 ties Self++ back to \emph{Dependent Origination}: the self is not fixed but co-arises with conditions, including tools, social relations, and institutions \citep{Macy1979DependentCoArising}. XR and AI become part of the causal web that shapes habits, identities, and norms; in turn, users and communities shape the objectives, feedback signals, and reward structures that shape AI systems. Read this way, Self++ is a co-evolutionary claim: \emph{self evolves, AI evolves, society evolves}. Without transparent reasons, adaptive sensitivity to context, and negotiable boundaries, the coupled system is prone to drift into manipulation, capture, or adversarial gaming. With the co-determination principles (T.A.N.), R9 aims at a constructive blurring of boundaries: not loss of agency, but authentic agency amplified, where the user becomes more competent, more autonomous, and more connected over time, and where the socio-technical ecology is steered towards those human ends rather than away from them.


\section{Self++ Design Propositions and Evaluation Checks}
\label{sec:selfpp_propositions}

Self++ is presented as an actionable framework, not only a taxonomy. In HCI and design-oriented research, frameworks become more reusable when they are articulated as explicit claims that others can inspect, debate, and evaluate across contexts, rather than only described narratively. This aligns with interaction-design arguments for making knowledge transferable through concrete representations and critique, accounts of intermediate-level knowledge that support reuse and cumulative learning across projects \citep{buxton2010sketching, hook2012strongconcepts, gaver2012annotatedportfolios}, and perspectives that emphasise articulating implications and principles so designs can be examined beyond a single instantiation \citep{dourish2006implications, carroll1992taskartifact, gregor2007anatomy}. It also aligns with Human-Centered AI arguments that safety, control, and responsibility must be operationalised as design requirements rather than stated only as values \citep{Shneiderman2020HCAI}.

Accordingly, we state a compact set of propositions that summarise what Self++ claims about human--AI coupling under SDT and FEP, and how to evaluate systems that aim to instantiate these claims. We present these propositions as falsifiable design hypotheses, not validated findings. Each states a necessary condition that Self++ predicts must hold for co-determined augmentation to succeed; Table \ref{tab:selfpp_propositions} pairs each proposition with concrete evaluation checks so that future empirical work can confirm, refine, or reject them. This distinguishes the propositions from heuristics or best-practice recommendations: they make specific, testable commitments about how SDT needs, FEP dynamics, and T.A.N. constraints interact under concurrent overlay activation, and they are intended to be wrong in instructive ways if the underlying theory is incomplete.

\begin{table*}[t!]
\centering
\scriptsize
\setlength{\tabcolsep}{4pt}
\renewcommand{\arraystretch}{1.25}
\caption{Self++ propositions (P1--P8) with brief evaluation checks. Use as a lightweight audit: map system features to Self++ role patterns, then verify T.A.N at the required strength and test transitions and drift under realistic overlap.}
\label{tab:selfpp_propositions}
    \begin{tabularx}{\textwidth}{@{}c >{\hsize=0.6\hsize}X >{\hsize=1.4\hsize}X@{}}
    \toprule
    \textbf{P} & \textbf{Proposition (what must be true)} & \textbf{Evaluation checks (what to test/measure)} \\
    \midrule

    P1 & \textbf{Concurrency:} Overlays act concurrently (not a pipeline) and can interfere. &
    Test overlap interference and recovery: (i) run Overlay 1 guidance while Overlay 2 deliberation UI is present (e.g., motor task + counterfactual dashboard) and measure errors/time-on-task; (ii) measure \emph{reclaim-time} (time to pause/override after an AI-led phase) and success rate of taking back control. \\
    
    P2 & \textbf{Timescale Alignment:} SDT needs map to uncertainty targets across temporal scales. &
    Evaluate on the right horizon: Overlay 1 with immediate sensorimotor metrics (errors, collisions, smoothness); Overlay 2 with decision quality \emph{and} goal-alignment/endorsement (regret, confidence, stated-goal match over days); Overlay 3 with longitudinal drift indicators (relationship repair, wellbeing, dependence, value-consistency) over weeks/months---not only short task scores. \\
    
    P3 & \textbf{Inspectability:} Legitimate augmentation requires an inspectable, contestable AI voice. &
    Probe legibility and ownership: users can say \emph{what} was influenced (evidence/salience/delegation), \emph{why}, and \emph{how to undo}. Include a behavioural test: can users find and use ``show alternatives/unnudged view'', ``undo'', and ``turn-off this support'' controls successfully? \\
    
    P4 & \textbf{T.A.N.\ Scaling:} Co-determination strength must scale with scope and initiative. &
    Audit \emph{T.A.N}: higher-scope, Overlay 3 (R7--R9) must provide (i) stronger reason/provenance and incentive disclosure, (ii) explicit consent gates and clearer boundaries, and (iii) easier reversal/audit trails than lower-scope, Overlay 1 (R1--R3). Example check: an R9 narrative nudge must offer ``see unframed view'' and ``disable this class of intervention''. \\
    
    P5 & \textbf{Transition Legibility:} Shifts in agency between role patterns must be perceptible and reversible. &
    Test hand-offs and failures (Advisor $\rightarrow$ Agentic Worker; escalation/de-escalation): users must correctly answer ``who is acting now?'', ``under what authority?'', ``how do I stop/revert?'' Measure situation awareness during transitions and recovery time/success after errors. \\
    
    P6 & \textbf{Endorsement over Compliance:} Autonomy support preserves endorsed priors, not compliance. &
    Check internalisation, not just performance: users endorse outcomes as \emph{their} choice and can explain ``because...'' in terms of their goals/values. Compare nudged vs unnudged conditions: if outcomes improve but endorsement drops or users cannot justify choices, autonomy support failed. \\
    
    P7 & \textbf{Collective Negotiability:} Relatedness support requires shared-model alignment and group negotiability. &
    Verify group legitimacy: collective opt-in for sensing/visualisations; privacy-by-role defaults; and opt-out without social penalty (no status loss, no exclusion cues). Test whether participants can contest aggregation rules/thresholds (e.g., participation metrics) and still collaborate smoothly. \\
    
    P8 & \textbf{Governance Contestability:} Long-horizon alignment is socio-technical and requires contestation pathways. &
    Audit ``contestability of action \emph{and} framing'': users can challenge recommendations \emph{and} the value assumptions/optimisation target behind them (e.g., ``stop treating this as productivity failure''). Verify governance hooks: escalation rules for intrusive interventions, audit logs, and pathways for collective contestation when groups/communities are affected. \\

    \bottomrule
    \end{tabularx}

\vspace{2pt}
\begin{minipage}{\textwidth}
\footnotesize
\textit{How to use (self-contained):} (1) Map features to Self++ role patterns (R1--R9) across the three concurrently activatable overlays (Table \ref{tab:selfpp_roles}). (2) For each claimed role pattern, verify co-determination (T.A.N.) commitments at the required strength (reasons/provenance/incentives; fading/calibration; override/contestability). (3) Evaluate transitions and long-horizon drift under realistic concurrent operation, not only steady-state task performance.

\end{minipage}
\end{table*}


\section{Exemplary Scenarios of Self++}

Meet Alex and Brooke, two 20-year-old university students facing the same three parallel demands: excelling in education, managing a part-time job, and maintaining a social life. Alex is steady and planful; Brooke is bursty and inconsistent. Brooke often stays up late gaming, wakes late, and misses classes, yet can become exceptionally creative and effective under pressure when they enter a flow state. Both adopt the Self++ XR system: an intelligent virtual assistant that runs mainly through \textit{XR glasses} running in AR mode in everyday routines and switches to VR for immersive practice, stress relief, or structured reflection.

Crucially, Self++ is not a linear ladder. Its three concurrently activatable overlays can combine Role Pattern R1-R9 under \textit{co-determination (T.A.N.)} (Transparency, Adaptivity, Negotiability), scaffolding growth while reducing uncertainty without taking ownership away from the user. This scenario uses a parallel timeline to show how the same nine role patterns can be adapted differently: for Alex, sustaining accumulation and calibration under pressure; for Brooke, enabling re-entry, preventing avoidable chaos, and preserving creative identity while building minimal, sustainable structure.

\subsection{Building Competence in Education (Tutor Mode)}

\textbf{Morning, 8:30 AM (Alex) / 11:30 AM (Brooke):} Alex heads to a chemistry lab for a new topic. Self++ enters \textbf{Tutor (R1)} and creates a safe, learnable corridor: directional cues, relevant equipment highlights, and step-gated safety procedures. As Alex measures chemicals, the system offers immediate, gentle corrections. The effect is twofold: early attributable success (SDT competence) and lower surprise (FEP), so anxiety drops.

Brooke wakes late, already behind, and is at risk of avoiding altogether. Self++ still uses \textit{R1}, but with a different aim: \textit{re-entry}. Instead of a full lesson, it compresses the task into the smallest viable corridor (``two actions only'') and reduces shame-driven uncertainty by making the next step unambiguous. If Brooke attends the lab after missing prior sessions, Tutor mode prioritises \textit{error prevention and safety gating} (what must not be missed) while keeping the interaction non-moralising and easily skippable. The goal is not discipline; it is enabling engagement in the first place.

\textbf{Afternoon, 2:00 PM (Alex) / 3:30 PM (Brooke):} A calculus assignment is due. Self++ shifts into \textbf{Skill Builder (R2)} and launches a VR practice module with an interactive whiteboard and immersive 3D visualisation. For Alex, it adapts difficulty on the fly and provides hints only after allowing time to think, keeping effort owned rather than outsourced. When Alex stalls, it uses subtle cueing (e.g., lightly highlighting a relevant formula) as a memory prompt rather than a solution dump. Support fades as proficiency stabilises. Alex experiences repeated, attributable successes, with challenge calibrated to avoid boredom or collapse into frustration.

For Brooke, R2 is structured as \emph{short, variable sprints} rather than long drills. The VR module reframes practice as micro-gamified challenges that preserve novelty while still training fundamentals. Hints remain optional and late, and the system schedules practice windows where Brooke is most likely to reach flow. The system builds competence by capitalising on Brooke's burst capacity, while quietly improving generalisation by varying contexts and constraints across sprints.

\textbf{Evening, 7:00 PM (Alex) / 11:30 PM (Brooke):} Approaching a mid-term test, Self++ becomes a \textbf{Coach (R3)}. For Alex, it overlays a heatmap on solutions (strong reasoning vs weak steps) and introduces metacognitive prompts. When it detects rushing through familiar sections, it nudges assumption checks. When Alex spirals after seeing peers post ``10-hour study days'', Self++ shows a private progress dashboard that grounds self-assessment in Alex's own trajectory rather than distorted social comparison. Coaching here trains resilience and calibration, preventing expertise from drifting into complacency or discouragement.

For Brooke, R3 targets a different brittleness: competence that appears mainly under adrenaline. Coach mode runs \textit{safe pressure practice} (timed scenarios, interruptions, missing information) and gives short debriefs focused on stabilising performance without extinguishing creative leaps. It adds a single guardrail against impulsive ``clever'' shortcuts (constraint checks) while protecting Brooke's ability to improvise. The point is robustness: creativity that remains reliable when conditions change, not just when panic peaks.

\subsection{Empowering Autonomy at Work (Advisor Mode)}

\textbf{Weekday, 9:00 AM (Alex) / 12:00 PM (Brooke):} Alex works part-time at a tech start-up. Self++ adopts \textbf{Choice Architect (R4)}: it shapes the decision context while preserving authorship. When Alex views the task board through AR, one or two tasks are gently highlighted because they match Alex's growth goals and the team's priorities. Alex can choose anything, but indecision costs less. Suggestions are transparently tagged as AI prompts, preventing the ``helpful layout'' from becoming invisible steering.

Brooke also benefits from R4, but the main risk is \textit{derailment by micro-choices}. Self++ makes ``tiny start'' actions the easiest to select (one visible tile that launches a 5-minute setup), and only adds friction where Brooke has explicitly opted in (e.g., a second confirmation before late-night gaming on weekdays). This preserves autonomy while reducing avoidable uncertainty created by impulsive context switches.

\textbf{Midday, 1:00 PM (Alex) / 2:30 PM (Brooke):} During a mixed-reality meeting, the team hits a snag. Self++ moves into \textbf{Advisor (R5)}. For Alex, it offers multiple options with brief justifications and visible uncertainties, rather than a single ``best'' answer. Alex contributes these as discussable alternatives, combining AI evidence with human judgement (team preferences, creative insight, organisational constraints). Trade-offs become legible rather than intimidating.

For Brooke, R5 is tuned for \textit{avoidance collapse}. The system uses concise counterfactuals and concrete next steps rather than long explanations. It may show two short futures (``if you delay'' vs ``if you do 20 minutes now'') and reframe tasks in Brooke's own value language (e.g., protecting creative identity by linking required work to personal projects). The system reduces decision entropy without converting into compliance pressure.

\textbf{Evening, 5:00 PM (Alex) / 6:30 PM (Brooke):} As Alex becomes more capable, Self++ supports \textbf{Agentic Worker (R6)}: delegated execution under a proposal-approval loop. Alex sets boundaries in an AR dashboard: draft reports automatically, but require review before sending; triage emails, but never touch messages marked sensitive. The system executes routine work quietly, pings for approvals at defined checkpoints, and stays interruptible. Autonomy strengthens because delegation is explicit, scoped, and revocable, and Alex learns meta-autonomy: when to hand off and when to stay hands-on.

For Brooke, R6 is ``anti-chaos delegation'': preventing administrative failure (missed emails, missed forms, missed replies) from consuming capacity and causing downstream social or institutional penalties. The system drafts messages and proposes schedules, but preserves consent checkpoints for anything consequential. Delegation here protects autonomy by preventing small failures from snowballing into externally imposed constraints.

\subsection{Fostering Relatedness in Social Life (Networker Mode)}

\textbf{Friday, 7:00 PM (Alex) / 8:30 PM (Brooke):} Alex and Brooke are friends, and they are both meeting the wider group at a café. Alex is keen but socially anxious, especially with new acquaintances, while Brooke arrives later and is noticeably quieter than usual. Self++ foregrounds Overlay 3 support for both of them. At \textbf{Contextual Interpreter (R7)}, the system reframes the evening as legitimate recovery rather than ``lost productivity'', showing Alex's completed commitments and a simple view of the week's balance. This reduces guilt and supports value-consistent wellbeing.

For Brooke, the challenge is often not anxiety but inconsistency: disappearing, then avoiding people due to embarrassment. R7 therefore makes consequences legible privately and without shame (e.g., ``you have not replied to X; a short repair message prevents drift''). It also clarifies social and institutional context (``this message expects a reply today'' vs ``FYI only''), reducing social surprise.

At the café, Self++ provides optional, privacy-respecting cues: names and agreed-to ``common ground'' hints for introductions. It stays light-touch: enough to reduce awkward uncertainty without making either user dependent. As the conversation unfolds, it shifts to \textbf{Social Facilitator (R8)}. When Alex notices Brooke is quiet, the system supports \emph{human-led inclusion} rather than stepping in as the social actor. In Alex's view, it offers a gentle, non-intrusive prompt such as ``Brooke has not spoken for a while; consider a check-in or an easy entry point'' and surfaces a low-stakes bridge topic grounded in shared context (e.g., ``ask about the design sprint they enjoyed''), without exposing private data. Alex uses this to invite Brooke in: a simple question, a shared joke, or an explicit acknowledgement (``glad you made it'') that lowers pressure.

For Brooke, R8 supports \emph{repair and re-entry} without public call-out. In their view, the system can offer opt-in micro-supports: a one-tap ``join-in'' suggestion (two or three possible responses), a private reminder that silence is acceptable, and a quick ``common ground'' cue that helps them re-engage on their own terms. If miscommunication appears, Self++ can offer a neutral micro-summary of viewpoints to prevent escalation. If the group is thriving, it does nothing. Relatedness is strengthened by improving human-human coordination and continuity, not by replacing relationships.

\textbf{Saturday, 10:00 AM (Alex) / 1:00 PM (Brooke):} The next day, Self++ runs a short \textbf{Purpose Amplifier (R9)} reflection. For Alex, in a calm AR ambience, it visualises how study, work, and social care link to longer-term aspirations. It reframes these as mutually reinforcing rather than competing: competence as foundation, autonomy as agency, relatedness as meaning and resilience. It may suggest value-aligned opportunities, but these remain invitations, not prescriptions. The purpose is coherence: reducing long-horizon drift by making values actionable without becoming coercive.

For Brooke, R9 protects creative identity while reducing drift that later feels like betrayal. The system does not prescribe ``be disciplined''; it supports value coherence through opt-in, inspectable simulations of downstream consequences (e.g., a ``future self'' contrast between chaotic nights and minimal structure that preserves creative time). It keeps framing legible and editable, ensuring Brooke can rewrite narratives in their own language.

\subsection{Balancing Conflicts via Co-Determination (where Self++ earns its keep)}

Life becomes convoluted when domains collide. During crunch week, Alex faces an exam, a critical work presentation, and a close friend's wedding within two days. Stress spikes because each demand threatens another.

\textbf{Anticipation and planning (weeks earlier):} Self++ notices the clash early and nudges forward preparation: earlier study blocks, a VR practice exam, and protected time around the wedding. This is active uncertainty regulation: fewer last-minute surprises means less stress.

\textbf{Negotiating autonomy (when work shifts):} Alex's boss asks to move the presentation to the wedding day. Self++ (\textit{R5--Advisor}) generates a private XR comparison of two timelines and their consequences. It finds feasible alternatives (another slot, coverage options) and helps draft a professional email proposing a solution. Alex remains the author; the system makes negotiation easier and less threatening. The boss agrees to reschedule.

\textbf{Dynamic rebalancing (day-of):} The exam and wedding still share a day. Self++ shifts roles fluidly: R2--Skill Builder/R3--Coach at dawn (focused VR review on weak areas), R4--Choice Architect/R5--Advisor before the event (logistics checks and timing nudges), R8--Social Facilitator at the wedding (mostly silent, with optional translation subtitles for an overseas relative), and recovery that night (a short VR calming session).

Brooke's conflicts often look different but are equally entangled: late-night flow collides with a Monday deadline, while a friend asks for help moving flat on Sunday morning. The system does not ``optimise'' Brooke; it makes the conflict legible and recoverable. A minimal co-determination response may combine: (i) \textit{R4} friction only where Brooke opted in (confirming the cost of starting another game), (ii) \textit{R5} two short futures (help friend + miss quiz vs delay help by 90 minutes and keep both), (iii) \textit{R8} a repair message draft (``I can help at 9:30; compulsory quiz at 8''), and (iv) \textit{R6} alarms and a checklist, with approvals at key points. Brooke still chooses; the system reduces avoidable surprise and supports agency-preserving recovery.

In both cases, success is not that Self++ ``won'' the trade-off, but that it helped keep all three SDT needs in view under pressure, while making interventions transparent, adaptive, and negotiable.

\subsection{Outcome: A Co-Determined Growth Trajectory}

Across domains, Self++ scaffolds without taking the steering wheel. In education, it builds competence from onboarding to robust mastery while training calibration: for Alex, steady accumulation and bias-resilient self-assessment; for Brooke, re-entry corridors, sprint practice, and pressure-safe robustness that protects creativity. At work, it strengthens autonomy from gentle prioritisation to explicit, reversible delegation: for Alex, throughput with oversight; for Brooke, anti-chaos delegation that prevents small failures from becoming externally imposed constraints. In social life, it reduces social uncertainty, supports repair, and deepens coherence with values and purpose: for Alex, confidence and presence; for Brooke, continuity and reconnection without shame.

When conflicts arise, the Overlays overlap rather than queue: competence support can run during autonomy negotiation inside a social obligation. Throughout, \textit{T.A.N} keeps augmentation legitimate: users can tell what is guided and why, support adapts and fades with growth, and overrides or renegotiations remain always available.

The result is not an overnight transformation but a sustainable trajectory: both users become more capable, more self-directed, and more connected, recovering quickly from mistakes because the system catches ``just enough'' to get back on track and then steps back.


\section{Discussion}\label{sec:discussion} 

Self++ responds to recurring XR--AI issues by treating \emph{co-determination} as a design requirement rather than a usability feature. Below, we consolidate the main implications into three themes: (i) agency and calibration, (ii) ethical boundary conditions for experience-shaping systems, and (iii) institutional and governance implications.

\subsection{Agency, calibration, and metacognitive accuracy}

A central risk in XR--AI assistance is \textbf{erosion of agency}: systems can take control ``for the user’s benefit,'' undermining learning, ownership, and accountability. Self++ counters this by keeping the human as the author of action while the AI scaffolds performance and decision-making. In \textit{T.A.N.} terms, \textit{Negotiability} operationalises consent, override, and renegotiation so assistance remains revocable and role boundaries stay explicit. This aligns with coactive teamwork, where human and AI remain interdependent partners rather than a controller and a controlled system \citep{johnson2011coactive}, and with mixed-initiative design that treats initiative shifts as coordination problems rather than hand-offs to be hidden \citep{Horvitz1999MixedInitiative}. A practical expectation is fewer mode-confusion episodes and fewer ``why did it do that?'' moments because intent and authority are made legible before the system acts.

A second challenge is \textbf{calibration}: users must calibrate both \textit{trust in the system} and \emph{confidence in themselves}. Poorly designed systems invite over-trust (misuse) or under-trust (disuse), undermining human--AI teaming \citep{lee2018trust}. XR can amplify these errors: immersive guidance can inflate perceived competence, while a single failure can collapse trust. Self++ addresses this through \textit{Transparency} and \textit{Adaptivity}: the system should disclose capability limits, intent, and uncertainty, and adjust autonomy as the user and context change. Clear uncertainty and rationale cues support appropriate verification \citep{okamura2020trustcalibration} and can improve satisfaction, situation awareness, and team performance \citep{yousefi2025team} by narrowing the ``gulf of evaluation'' between user expectations and system behaviour \citep{norman1988psychology}.

Self++ also treats \textbf{self-assessment biases} as part of calibration. Novices can overestimate mastery while experts underestimate gaps; XR training that maximises ease can worsen these illusions by confounding performance with assistance. Self++ therefore emphasises calibrated feedback, scaffolded reflection, and guidance fading: the system should make the source of success legible (user skill vs. AI help) and progressively withdraw support as competence stabilises. This aligns with evidence on prompting explanation and fading hints \citep{anderson1995cognitive}, and with ``desirable difficulty'' accounts showing that structured challenge improves retention and reveals limits \citep{bjork2011making, landman2018training}. More broadly, immersive representations can be used for reflective sensemaking when they externalise uncertainty structures and alternatives rather than presenting a single persuasive conclusion \citep{Davidson2023Sensemaking}. When paired with plural perspectives in deliberation, these mechanisms can reduce cognitive illusions amplified by digital mediation \citep{fisher2015searching, sparrow2011google}.

\subsection{Ethical boundary conditions for experience-shaping systems}

Because XR systems can shape the evidential stream, Self++ is not a moral optimiser and should not be framed as ``making people good.'' Its goal is to help users act more consistently with what they already endorse, while keeping influence inspectable and revisable under T.A.N. This stance requires an explicit acknowledgement: Self++ is ethically procedural, not substantive. It does not encode a preferred moral framework or assume universal agreement on what constitutes a good life, a responsible choice, or a well-functioning community. Such standards vary across cultures, belief systems, and individual histories, and any system that hard-codes a single ethical orientation risks becoming an instrument of cultural imposition rather than support. Self++ addresses this by locating its normative commitments at the level of process rather than content. T.A.N. specifies how the system must behave, making influence visible, tuning support to context, and preserving the user's right to contest and override, without specifying what the user should value or endorse. 

However, this procedural stance is not value-free. The decision to surface certain consequences rather than others, to frame trade-offs in particular ways, or to define what constitutes ``drift'' from endorsed values all involve normative assumptions that may reflect the designers' cultural position. Self++ therefore treats these assumptions as first-class design parameters: they must be documented, auditable, and revisable through the negotiability mechanisms described in Sections \ref{sec:overlay3}, and through the participatory and co-design approaches noted in Section \ref{sec:limitations-future-work}. Where systems affect communities rather than individuals alone, collective contestation pathways (R8's collective negotiability and R9's governance hooks) become the primary safeguard against monocultural default assumptions. In short, Self++ aspires to cultural humility: it holds that the conditions for flourishing are real and important, but that their specific expression must be determined with users and communities, not for them. This distinction becomes especially important in higher-scope interventions (identity, relationships, civic judgement, long-horizon behaviour), where even well-intentioned support can become covert preference shaping.

Self++ also has a clear \textbf{dual-use} risk. The same mechanisms that scaffold autonomy and relatedness can be repurposed as manipulation, including persuasive dark patterns \citep{luguri2021darkpatterns, mathur2019darkpatterns} or ``sludge'' that preserves the appearance of choice while steering outcomes \citep{schmidt2020ethics}. Personalisation data can further enable profiling and undue influence. At minimum, safeguards should follow ethical nudging guidance \citep{meske2020ethicalnudges} and human-centred AI principles \citep{Shneiderman2020HCAI}: transparency should include \emph{intent} (who benefits), users should have meaningful opt-out and data control, and systems should be auditable for manipulative behaviour.

XR introduces an additional manipulation surface via \textbf{self-presentation}. Systems that filter or reframe a user’s social signals (e.g., making them appear happier or suppressing negative affect) can function as social dark patterns if users cannot inspect or contest the transformation \citep{Hart2021Avatars}. Even when users consent initially, default-on transforms risk identity drift and misattribution in consequential settings (work evaluation, conflict repair, health, legal contexts). A Self++-consistent constraint is: self-presentation interventions require high-salience disclosure, editable parameters, and easy reversion, with stronger safeguards as stakes rise.

A related policy gap concerns \textbf{state-aware assistance} when the user is plausibly impaired (drowsy, medicated, intoxicated, acutely stressed). State sensing can increase safety, but it also increases surveillance and paternalism risk. A Self++ pattern is to treat impairment detection as \emph{risk gating}, not permission to seize control: disclose what is sensed and its reliability, shift to safer defaults (more confirmations, reduced autonomy, fewer irreversible actions), and require explicit, revocable opt-in for any escalation beyond nudges. In group settings, inferring impairment is sensitive, so sharing it outward should be prohibited by default except for clearly defined, consented safety protocols.

Finally, the next generation of XR will increasingly generate \emph{experience} (adaptive soundscapes, affective ambience, personalised visuals, fully generative one-off environments). These can support restoration, creativity, and engagement, but also introduce covert mood steering and narrative capture. Self++ therefore treats \textbf{framing legibility} as a hard requirement: users must be able to distinguish evidence from aesthetic framing and persuasive scaffolding, and adjust or disable these overlays. This also applies to wellbeing applications such as mindfulness and self-transcendent experiences, which are promising but high-leverage; reviews suggest XR can support contemplative practice when interventions are bounded and autonomy-supportive \citep{kitson2020review}. Here, the T.A.N gradient matters: intent disclosure, adjustable intensity/frequency, and debrief mechanisms help users integrate benefits without dependency.

On the theme of \textbf{transcendence}, this ambition resonates with a Buddhist view that liberation becomes possible through insight into how experience is conditioned, classically articulated through dependent origination \citep{Macy1979DependentCoArising}. In this account, ignorance is not a lack of knowledge but a structural error in perception: treating impermanent, interdependent processes as fixed and self-contained, including the construct of a stable, separate self \citep{gallagher2024self}. Contemplative practice aims to correct this error by making the causal chains between perception, craving, and habitual reaction visible and interruptible. Self++ shares this structural logic without claiming equivalence: by making the conditions of mediated experience transparent, adaptive, and negotiable, the framework supports the user's capacity to notice what is being shaped, by whom, and toward what ends. In this reading, co-determined XR-AI systems could function as attentional scaffolds that sustain the reflective clarity that contemplative traditions regard as a prerequisite to wise action, provided such systems remain bounded, autonomy-supportive, and subject to the user's ongoing consent \citep{kitson2020review}.

\subsection{Beyond Human-Level Reasoning: Self++ as an Interface for Superhuman and Self-Improving AI}

A further motivation for Self++ is the plausible trajectory toward \textbf{artificial super intelligence (ASI)}\citep{Bostrom2014Superintelligence}, including systems that improve via self-play, self-generated curricula, recursive self-improvement, or scalable oversight beyond direct human feedback. AlphaGo highlighted how learned policies can produce strategies that surprise experts, and AlphaGo Zero strengthened the point by reaching high performance with minimal human priors beyond the rules \citep{Silver2017GoZero}. In broader optimisation and scientific settings, analogous agents may propose solutions in high-dimensional spaces that are useful yet difficult for humans to justify or even interpret. This creates an \emph{interface problem} as much as a capability problem: when reasoning outruns ordinary human intelligibility, the risk is not only power, but loss of the ability to \emph{understand, contest, and appropriately rely on} proposals, a concern central in control and alignment work \citep{Bostrom2014Superintelligence, Russell2019HumanCompatible, Amodei2016Concrete}. XR can exacerbate misplaced deference because immersive presentation can add authority while hiding assumptions and failure modes; however, it can also be part of the remedy by using human multimodal perception for higher-bandwidth sensemaking (spatialised structure, audio, haptics, thermal cues) so users can more fully \emph{grok} complex trade-offs in parallel. Recent work on \emph{superalignment} frames this as a weak-to-strong governance problem: humans are the weak supervisors who must still control systems much smarter than themselves, so interfaces like Self++ become part of the oversight stack that keeps decisions inspectable and contestable rather than merely convenient \citep{Burns2023WeakToStrong}.

Reframed in Self++ terms, superhuman reasoning raises the required strength of \textit{T.A.N} rather than weakening it. \textit{Transparency} must shift from “explain the answer” to “make the \emph{decision structure} legible”: expose constraints, trade-offs, counterfactuals, and uncertainty in forms people can interrogate, aligning with interpretability aims of human-meaningful representations \citep{Olah2017FeatureVis, Olah2018BuildingBlocks}. \textit{Adaptivity} must tune that legibility to human limits and stakes (what to surface now, what to defer, when to escalate evidence), while considering epistemic humility about boundary conditions and distribution shift \citep{Russell2019HumanCompatible}. \textit{Negotiability} becomes the core safety valve under asymmetric intelligence: even if the system can discover options humans would not find, \emph{adoption remains co-determined} via explicit veto points, staged commitments, and contestable assumptions, echoing the motivation for scalable supervision and preference-based oversight while recognising their limits \citep{Christiano2017HumanPreferences, Amodei2016Concrete}. In this reading, Self++ treats XR as a sensemaking overlay between human values and superhuman optimisation: advanced intelligence can be usable without becoming unquestionable, because T.A.N keeps proposals inspectable, adjustable to context, and always contestable under human authority.

\subsection{Limitations and future work} \label{sec:limitations-future-work}

Self++ is a role-based interaction theory, so its main limitations are less about conceptual coverage and more about \emph{operationalisation}: building systems that deliver co-determined support reliably, measuring SDT-relevant states in situ, and validating effects over time and across contexts.

\textbf{Operational feasibility in real-world XR.} Running multiple roles as concurrently activatable overlays requires real-time policy arbitration, conflict handling, and fast failure recovery, and making automation behave as a true team player remains demanding in practice \citep{seeber2020machines, klein2004teamplayer}. Many interactions also assume robust sensing and timely feedback; current hardware constraints can break legibility cues or mis-trigger interventions, and response delays can degrade trust, coordination, and perceived social presence \citep{templeton2022fastresponse}. A practical agenda is to specify role-specific tolerances (latency, sensing fidelity), then design graceful degradation paths when those tolerances are not met.

\textbf{Interaction design of role-pattern transitions.} Self++ specifies what should change when the system shifts between role patterns, the functional intent, support level, and T.A.N. requirements, but deliberately leaves underspecified how that change is communicated to the user in XR. When the system transitions from Tutor (R1) to Skill Builder (R2), or when Coach (R3) and Advisor (R5) activate concurrently during a team training scenario, the perceptual and interaction design of that transition, whether it manifests as a gradual fading of visual cues, an explicit notification, an ambient shift in soundscape or colour temperature, or a change in agent embodiment or behaviour, remains an open design research question. This omission is intentional: the appropriate transition idiom is likely to be highly dependent on modality (AR vs VR), task criticality, attentional capacity, and user preference, making premature specification counterproductive. However, that transition design is not merely cosmetic. Poorly communicated role shifts risk the mode confusion and automation surprise that Self++ aims to prevent (Section \ref{sec:r3-coach}), while overly salient transitions may disrupt flow or impose unnecessary cognitive load. We therefore invite empirical investigation into transition legibility, including comparative studies of implicit (ambient) versus explicit (announced) role-shift cues, user-configurable transition salience, and the perceptual markers that best support situation awareness during concurrent overlay activation. Table \ref{tab:selfpp_propositions}, proposition \textit{P5}, provides initial evaluation criteria for this work.

\textbf{Measurement, legibility, and the cost of co-determination.} Self++ presumes systems can tune support to competence, autonomy, and relatedness dynamics, yet reliable real-time indicators for these constructs remain limited. Trust and reliance have workable behavioural signals (for example, hesitation and overrides) \citep{wischnewski2023trust}, but analogous indicators for competence frustration or relatedness quality are underdeveloped. Future work should develop lightweight in situ measures (micro-self-reports and unobtrusive multimodal signals) that are accurate enough to drive adaptation without becoming intrusive or surveillance-like. In parallel, designers must avoid over-scaffolding: persistent support can create dependency and out-of-the-loop problems \citep{endsley1995outoftheloop}, and can inflate self-assessment when assistance is confounded with skill \citep{kruger1999unskilled}. 

While guidance fading and structured challenge offer principled countermeasures \citep{atkinson2000studying, bjork2011making}, the right fade schedule is task- and person-dependent, so scaffolding schedules should be treated as testable design parameters, including whether calibration cues reduce inappropriate deference \citep{okamura2020trustcalibration}. Legibility also becomes harder at higher overlays: explaining context interpretation or long-horizon coaching requires user-meaningful and auditable accounts, but much XAI work targets local model decisions rather than extended cognitive interventions \citep{haque2023xaiuserperspective}. XR further constrains explanation because it must fit attention limits without breaking immersion; early AR-focused frameworks are promising but immature \citep{xu2023xair}. Finally, transparency and negotiability introduce overhead: over-disclosure and frequent confirmations can interrupt flow and increase workload \citep{buchner2022impact}, echoing classic concerns that excessive mode signalling can harm situation awareness \citep{sarter1995mode}. Future systems should therefore prioritise selective, stake-triggered transparency, quiet background cues, and user-tunable thresholds.

To reduce the gap between theoretical constructs and engineering implementation, a practical next step is to translate the co-determination principles into concrete programmable logic. For Negotiability, systems could incorporate a “Trust Protocol” built around pre-assigned autonomy budgets or risk thresholds. This would allow roles such as the Agentic Worker to execute low-stakes actions without constant confirmation, while reserving interruptions for exceptions and boundary conditions. Likewise, the timing of Adaptivity could be formalised via “Drift Triggers”: statistical thresholds (for example, a plateau in performance variance) that indicate reduced entropy and automatically prompt a transition from support (Skill Builder) to challenge (Coach). Finally, validation will require evolving the evaluation checks (Table \ref{tab:selfpp_propositions}) from qualitative audits into quantitative proxy metrics, such as tracking reclaim rates (the frequency of user overrides) to infer autonomy calibration, or measuring turn-taking balance to quantify relatedness quality in multi-agent settings. To be clear, the propositions in Table \ref{tab:selfpp_propositions} are offered as empirically testable claims, not as design guidelines whose validity is assumed; progress requires that future studies treat negative or partial results as informative refinements of the framework rather than implementation failures. 

\textbf{Generalisability, integration, and evaluation infrastructure.} Relatedness and autonomy are expressed differently across cultures and contexts, so Self++ needs stronger guidance on how interaction styles and boundaries should vary under different self-construals and relational norms \citep{markus2014culture, wilson2021creating}. Because Self++ touches identity-, relationship-, and purpose-adjacent support, participatory and co-design approaches are important, particularly with marginalised groups who may face distinct risks and expectations \citep{dourish2006implications}. Implementation choices will also be shaped by AI capabilities: large multimodal models could expand context understanding, dialogue, and planning \citep{yang2025magma}, but raise controllability and presentation challenges, including how to surface uncertainty, provenance, and constraints in user-legible form \citep{liao2023ai}; one pragmatic direction is hybrid architectures where foundation models are sandboxed or used offline, while real-time interaction is governed by stricter policies aligned with T.A.N. Finally, progress will be slow without shared benchmarks and compliance metrics: beyond outcome measures, we need longitudinal and teaming-quality evaluations, including when combined human--AI systems outperform human-only or AI-only baselines \citep{vaccaro2024combinations}, operational measures of team fluency \citep{mathieu2000influence}, domain norms for organisational settings \citep{Herrmann2022socio}, and explicit metrics for T.A.N compliance suitable for audit and comparison. These efforts align with human-centred AI guidance \citep{Shneiderman2020HCAI} and emerging trustworthy agent frameworks \citep{cheng2026safeagents}, and could provide a foundation for reproducibility and eventual standardisation.

\section{Conclusion}\label{sec:conclusion} 

Self++ advances a theory of human–AI teaming for XR that treats “help” as a coupled relationship rather than a one-way service. It starts from the premise that effective augmentation must grow the person, not quietly replace them. Grounded in basic psychological needs from Self-Determination Theory (autonomy, competence, relatedness) and the Free Energy Principle’s emphasis on stability under uncertainty in perception and action, Self++ frames good assistance as support that remains contestable, adjustable, and accountable.

The framework makes this actionable by organising augmentation into three interlocking overlays: Self for sensorimotor competence support, Self+ for deliberation and choice support, and Self++ for social, identity, and long-horizon alignment. These overlays are not a maturity ladder but overlays that can be activated as the situation demands. Across them, Self++ articulates role-based patterns (rather than anthropomorphic personas) and an interactional stance that keeps intent, limits, and uncertainty legible, so users can meaningfully endorse or refuse the system’s contributions.

Ultimately, Self++ is a blueprint for a symbiotic cognitive niche in the spirit of J. C. R. Licklider’s vision of tight human–computer partnership and the “coupled system” perspective in Andy Clark and David Chalmers. In this niche, the human supplies purpose, values, and accountable will, while the AI supplies navigable pathways, options, and scaffolding. The future is neither automated nor purely human-led, but co-determined through interactions designed to preserve agency while extending what people can perceive, decide, and become.

\newpage
\printbibliography

\end{document}